\documentclass[11pt]{article}
\usepackage[a4paper, total={6in, 9in}]{geometry}
\usepackage[utf8]{inputenc}
\usepackage{color}
\usepackage{soul}
\usepackage[dvipsnames]{xcolor}
\usepackage{authblk}
\usepackage{graphicx}

\usepackage[backend=biber,style=nature]{biblatex}
\usepackage{makecell}
\usepackage{pifont}
\usepackage{lineno}
\usepackage[font=scriptsize,labelfont=bf]{caption}

\title{Dynamic stiffening of the flagellar hook}
\author[1]{Ashley L. Nord}
\author[1]{Ana\"{i}s Biquet-Bisquert}
\author[1]{Manouk Abkarian}
\author[2]{Théo Pigaglio}
\author[2]{Farida Seduk}
\author[2]{Axel Magalon}
\author[1]{Francesco Pedaci}

\affil[1]{Centre de Biologie Structurale, CNRS, INSERM, Univ. Montpellier, France}
\affil[2]{Aix Marseille Université, CNRS, Laboratoire de Chimie Bactérienne (UMR7283), IMM, IM2B, 13402 Marseille, France}
\date{}

\begin{document}
\maketitle

\textbf{Many bacteria are motile by means of one or more rotating rigid helical flagella, making them the only known organism to use rotation as a means of propulsion. The rotation is supplied by the bacterial flagellar motor, a particularly powerful rotary molecular machine. At the base of each flagellum is the hook, a soft helical polymer that acts as a universal joint, coupling rotation of the rigid membrane-spanning rotor to rotation of the rigid extra-cellular flagellum. 
In multi-flagellated bacterial species, where thrust is provided by a hydrodynamically coordinated bundle of flagella, the flexibility of the hook is particularly crucial, as many of the flagella within the bundle rotate significantly off-axis from their motor. But, consequently, the thrust produced by a single rotating flagellum applies a significant bending moment to the hook. So, the hook needs to simultaneously provide the compliance necessary for off-axis bundle formation and the rigidity necessary to withstand the large hydrodynamical forces of swimming. 
To elucidate how the hook can fulfill this double functionality, measurements of the mechanical behavior of individual hooks under dynamical conditions are needed. Here, via new high-resolution measurements and a novel analysis of hook fluctuations during \textit{in vivo} motor rotation in bead assays, we resolve the elastic response of single hooks under increasing torsional stress, revealing a clear dynamic increase in their bending stiffness. Accordingly, the persistence length of the hook increases by more than one order of magnitude with applied torque. Such strain-stiffening 
allows the system to be flexible when needed yet reduce deformation under high loads, allowing cellular motility at high speed.}
\vspace{1cm}

Many soft biological materials exhibit a non-linear elastic behavior wherein they become stiffer upon deformation \cite{storm2005}. Such strain-stiffening behavior arises in a diverse collection of biological materials and over a variety of scales, with examples including the fibrin gels responsible for blood clotting \cite{janmey1983}, actin filaments within cellular cytoskeletons \cite{gardel2008}, corneal tissue \cite{elsheikh2007}, the walls of blood vessels \cite{ohayon2011}, the collagen fibers of tendons \cite{miller2012}, and the extracellular matrix of the lung \cite{jorba2019}. While such an elastic response often serves a critical physiological role, such as preventing  damage upon exposure to large deformations, the molecular and structural design principles which underpin the behavior remain largely unknown.
An exquisite example of fine-tuning of mechanical properties in biological systems is the rotating flagellum which provides the thrust required for many motile bacteria to move about and explore their environment.  The rotation is supplied by the bacterial flagellar motor (BFM), a large and powerful rotary engine spanning the bacterial membranes. The rotation of the cytoplasmic rotor is coupled to the rod, the central drive shaft which spans the membranes. The rod imposes a moment normal to the cell surface which is transmitted via the hook, a short ($55-60$ nm) extracellular polymer filament, to the microns long flagellar filament \cite{hirano1994, waters2007flik}. All three of these components are helical, hollow, slender rods, and the proteins which compose them (FlgG, FlgE, and FliC, for the distal rod, hook, and flagellum, respectively) all share high sequence homology with similar quaternary structure and helical symmetry. Yet, the three structures have strikingly different mechanical properties. The rod is straight and rigid, the hook is supercoiled and flexible, and the flagellum is supercoiled and $\sim$ 2 orders of magnitude more rigid than the hook \cite{fujii2017identical, johnson2021molecular, tan2021structural, shibata2019torque, kato2019structure, son2013bacteria}. Together, they provide an intriguing example of how similar sequences and structural motifs can beget largely distinct mechanical properties.  

Rotating at speeds up to hundreds of Hertz and propelling the cell at tens of microns per second, the flagellar filament is subject to high hydrodynamical loads, and its integrity relies upon its rigidity. 
It has been shown both theoretically and experimentally that flagellar bundling is impossible if the hook is not sufficiently flexible \cite{brown2012flagellar, lee2018, riley2018swimming, nguyen2018}. Moreover, the length of the hook is tightly controlled; mutations which shorten the hook length, thereby increasing the bending stiffness, disrupt the universal joint function, whereas longer hooks lead to instability in the flagellar bundle and impaired motility \cite{sporing2018}.
Thus, bacterial motility relies upon a delicate combination of flexibility and rigidity in a single appendage. Previous work has shown that the relaxed hook is so flexible that it can buckle under compression, an instability that monotrichous bacteria exploit to reorient their swimming direction \cite{son2013bacteria}. This effect is likely even more important in peritrichous bacteria, where the thrust of the off-axis flagellum applies a bending moment to the hook.  So, how is the hook able to be soft enough to enable bundle formation yet rigid enough to withstand the force of the rotating flagellum? 
In mono-flagellated polar bacteria, it has been proposed that the hook (subject to axial stress) becomes stiffer under increasing load \cite{son2013bacteria}. However, the mechanical response of a single hook to changing conditions is lacking, and the current knowledge of hook mechanics relies on sparse data and population averages. 

Here, measuring at high spatio-temporal resolution (sub-ms, $\sim 10$ nm) the position of a micro-bead tethered to a functioning motor, we develop a novel fluctuation analysis which allows to dynamically quantify the bending stiffness of single hooks in multi-flagellated (peritrichous) \textit{E. coli}. We observe the evolution of the hook bending stiffness in individual motors as they transition from zero speed to full physiological speed over a time scale of minutes. Our measurements show a clear dynamic stiffening of the hook as a function of motor speed, which scales with the imposed twist. We thus provide quantitative, \textit{in-vivo}, and time-resolved evidence for dynamic torsional-strain induced stiffening of the bacterial hook, adding this universal joint to the list of strain-stiffening biopolymers.  
This mechanical phenomenon may prove to be widespread among bacteria, despite different mechanisms of motility and differences in composition and sequence length of the hook protein \cite{yoon2016}.

\begin{figure}
\centering
\includegraphics[width=0.88\linewidth]{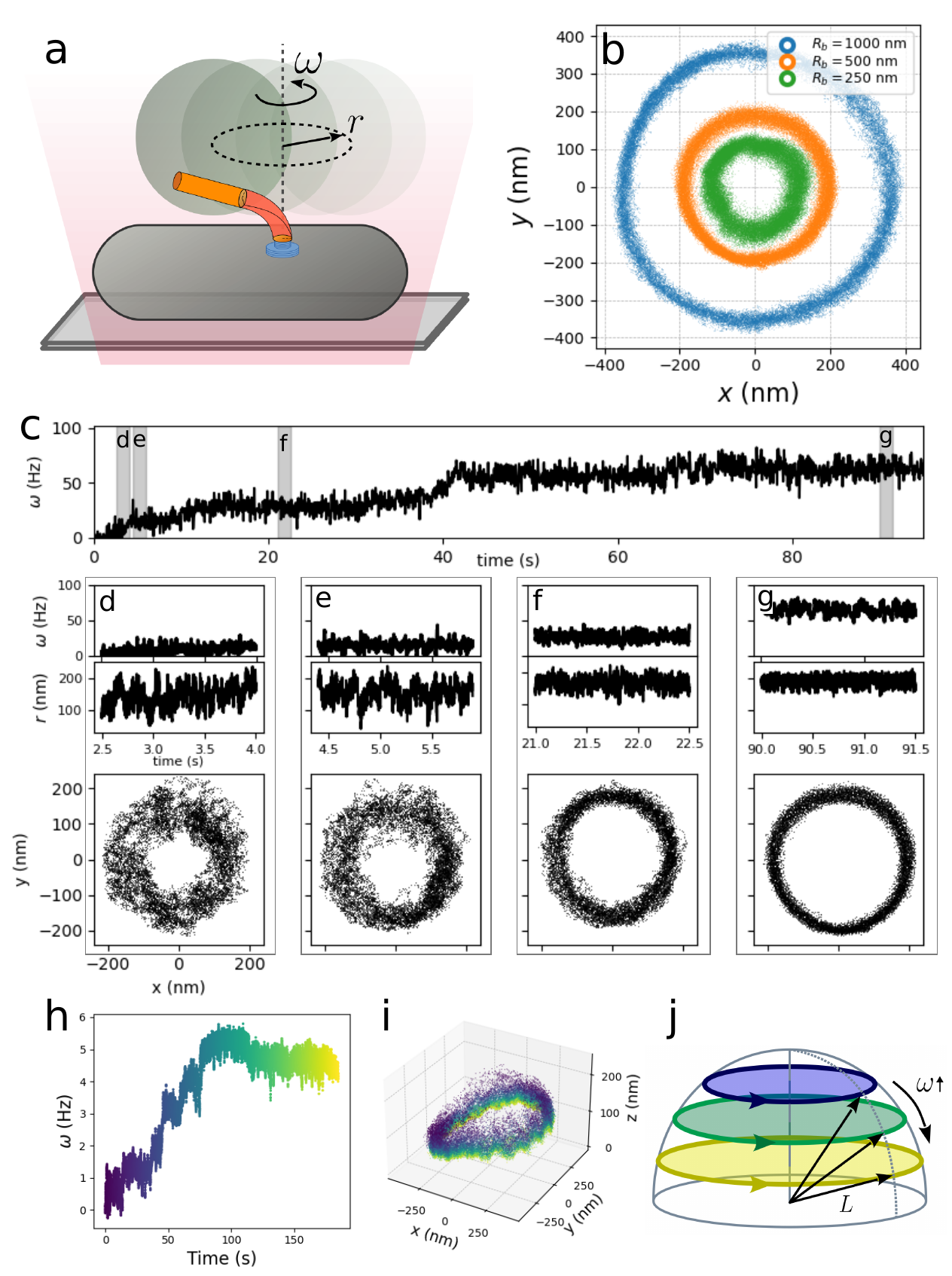}
\caption{Bacterial Flagellar Motor bead assays. 
a) Schematic experimental setup (not to scale). A living bacterium is adhered to the microscope glass slide, and one microscopic bead is attached to the filament stub, rotating on an approximately circular trajectory, observed by optical microscopy and tracked at an 8-10 kHz sampling rate.
b) Experimental trajectories of the bead center obtained for beads of different diameter.
c) Resurrection trace ($R_b=500$ nm). Along the time trace of the motor speed $\omega(t)$, four time-windows are highlighted, and for each window the corresponding signals $\omega(t)$, radius $r(t)$, and the bead trajectory $x(t),y(t)$ are shown in the respective panels (d-g). The decrease in radial fluctuations at increasing speed is particularly evident.
h) A resurrection trace (from a different motor than in c), where the color code indicates the time. 
i) 3D tracked position of the bead corresponding to the measurement in h), with  the same color code for time. The bead starts far from the membrane for low $\omega$ and approaches it when rotating faster, tracing circles of larger diameter. The trajectory is tilted because of the presence of the cell body. 
j) Schematic representation of the behavior of the trajectory in i). We simplify the geometry assuming that the bead center moves on circular trajectories on a hemispherical surface of fixed radius $L$. The color code indicates time as in h) and i).}
\label{fig_1}
\end{figure}
\section*{Results}

\subsection*{Radial fluctuations in bead assays}

Widely used to study the dynamics of individual BFMs, bead assays allow fast dynamical measurements of the motor activity in controlled conditions, including over a wide range of loads and external manipulation \cite{sowa2008bacterial, nord2020mechanisms}.
In bead assays, a microscopic bead of known diameter is tethered to a flagellum of a living cell adhered to the microscope slide (Fig.\ref{fig_1}a).
When successfully tethered to a single motor, the bead rotates slightly off axis, a clear signature of the activity of the motor (Fig.\ref{fig_1}b). 
By fitting an ellipse to the tracked bead trajectory (projected onto the $xy$ plane of the image), it is possible to assign an angle to each point, which reflects the angular displacement of the motor, assuming that the bead spin and orbital speeds are equal (i.e. that the bead rotates, as the Moon around the Earth, keeping the same face towards the center).
In the past decades, many results about the BFM inner mechanisms have been obtained via bead assays, utilizing the \textit{tangential} displacement of the bead. However, the microscopic description of the system still remains vague, and key features such as the geometry of the tether, the distance between the bead and the membrane, the bead viscous drag, and as a consequence, the value of the torque produced, remain to be elucidated. 
Here we show that, together with the tangential movement, the \textit{radial} displacement of the bead along its trajectory reports upon the mechanical properties of the hook's universal joint mechanism. In particular, we observe that the bead's radial fluctuations decrease with increasing motor speed. To explain such behavior, we constructed a simplified geometrical model of the system, compatible with our observations, wherein the radial movement of the bead corresponds to the bending of the hook. Together with the hook bending stiffness, our measurements and model allow us to estimate the distance between the bead and cell membrane, which allows us to calculate the membrane's contribution to the drag coefficients of the bead, and to accurately measure the torque produced by the motor.

To change motor speed, we performed resurrection experiments \cite{block1984successive} on a non-switching strain of \textit{E. coli} which expresses a hydrophobic filament (FliC$^{sticky}$) that binds to polystyrene micro-beads after mechanical shearing. 
In Fig.\ref{fig_1}c we show a resurrection measurement where we observe a clear change of radial fluctuations in the $x,y$ bead trajectory. With increasing motor speed $\omega$, the radial fluctuations decrease, while the average radius of the trajectory increases, as outlined in four time-windows along the trace. 
Tracking beads in 3 dimensions, we observe that the bead center explores an approximately hemispherical surface while rotating, starting far from the membrane at low speed, and ending on a larger circle closer to the surface at higher $\omega$ (Fig.\ref{fig_1}h-i, similar measurements are shown in the SI). 
We summarize these observations in Fig.\ref{fig_1}j, which sketches the geometry we define in more detail below.

\subsection*{Geometrical model}
In Fig.\ref{fig_2}a we outline a plausible simplified geometry of the bead tethered to the flagellar stub, composed by the hook and the truncated FliC$^{sticky}$ portion. 
The tangential variable commonly tracked in bead assays is the angle $\phi$, together with its time derivative $\omega$ which corresponds to the angular motor speed when the bead (of radius $R_b$), once hindered by the surface, cannot spin around the tilted flagellar axis but can only rotate around the vertical motor axis $z$. With smaller beads, or beads attached at a larger radius along the flagellar stub, this geometrical constraint can be relaxed, leading to the observation of precession during rotation \cite{shimogonya2015torque}.
The radius $r$ indicates the distance of the bead center from the center of the $x,y$  trajectory (example time traces $r(t)$ can be seen in Fig.\ref{fig_1}d-g).
In the ($z,r$) plane (Fig.\ref{fig_2}b), $L$ is the distance between the bead center and the origin (the point where the hook intersects the membrane) and $\theta$ is the angle between $L$ and the membrane, which we approximate for simplicity as a plane.
We assume that the hook is a soft angular spring \cite{son2013bacteria, shibata2019torque}, and that $L$ is constant due to the negligible extension of the hook and filament, so changes of the radius $r$ and angle $\theta$ reflect the bending of the hook (as also suggested in \cite{wang2020measurement}). 
In this way, the measured probability distribution of $r$ (of width $\sigma_r$, Fig.\ref{fig_2}b) reflects the visited angles $\theta$.
The gap $s$ between the bead and the membrane is not directly measured (the bead position is tracked relative to an arbitrary origin), and changes in time with $\theta$ and $r$. Its minimum value for the entire trajectory is left as a free parameter, and its value $s_{\mbox{\tiny min}}$ is determined by the analysis described below. 
With our assumptions, $s_{\mbox{\tiny min}}$ corresponds to the extreme radial value $r_{\mbox{\tiny max}}$ visited in the entire  trajectory (Fig.\ref{fig_2}b).
Therefore, our assumptions can be summarized as follows: at a given time $t$ (Fig.\ref{fig_2}c), the bead is observed at a position $x(t),y(t),z(t)$, at a constant distance $L$ from the motor, described by a radius $r(t)$ and angle $\theta(t)$, corresponding to a bead-membrane gap of $s(t)$. Our analysis focuses in particular on the fluctuations of the angle $\theta(t)$.

\subsection*{Drag coefficients}
\begin{figure}
\centering
\includegraphics[width=.75\linewidth]{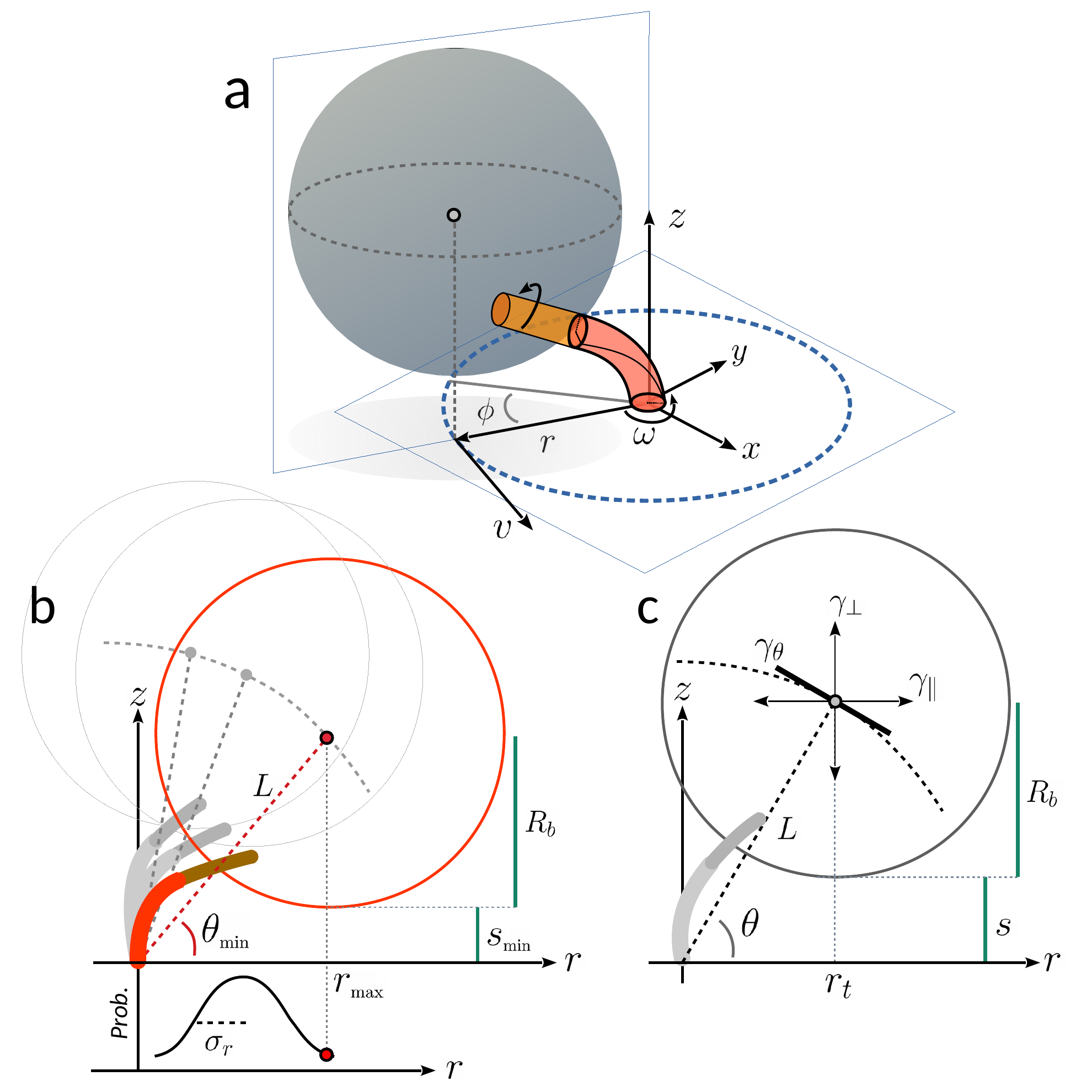}
\caption{Microscopic geometrical model. 
a) 3D representation of the bead (not to scale) tethered to the flagellar stub, composed by the hook (red, length of 60 nm) and filament (FliC$^{st}$, orange, of 20 nm diameter \cite{namba1989structure}). 
The angle $\phi$ describes the motion of the center of the bead along the circular trajectory of radius $r$. The linear speed of the bead is $v=\omega r$, where the angular velocity is $\omega=d\phi/dt.$
b) Projection on the ($z,r$) plane, where the center of the bead is assumed to move on an arc of radius $L$, described by the angle $\theta$. The probability distribution of $r$ (approximated by a Gaussian of width $\sigma_r$) is shown in the graph below. The minimum value of $\theta$ visited in an entire measurement ($\theta_{\mbox{\tiny min}}$) corresponds to the maximum visited value of the radius $r_{\mbox{\tiny max}}$, and to the minimum distance $s_{\mbox{\tiny min}}$ between the bead surface and the membrane.
c) Same as in b) for a generic position ($r_t,\theta$) at time $t$. The drag coefficient on the plane $(z,r)$ is composed by a parallel ($\gamma_\parallel$) and perpendicular ($\gamma_{\perp}$) component with respect to the membrane, which are projected ($\gamma_{\theta}$) on the direction tangent to the arc of radius $L$. The distance between the bead surface and the membrane is $s(t)\ge s_{\mbox{\tiny min}}$. The radius of the bead is $R_b$.
}
\label{fig_2}
\end{figure}

In this overdamped system, the proximity of the bead to the cell membrane must be taken into account to correctly quantify the viscous drag acting on the bead. 
An accurate value of the drag is required to determine the value of motor torque.
For a displacement parallel to the $(x,y)$ plane in the direction of the tangential linear speed $v$ (Fig.\ref{fig_2}a), we calculate the drag $\gamma_\phi$ (function of $s,R_b,\langle r \rangle$) employing both Brenner and Faxen theory, depending on the ratio $s/R_b$, to correct for the vicinity of the surface \cite{brenner1961slow, leach2009comparison, chio2020hindered}, as detailed in the SI.
The motor torque can then be calculated by $\tau=\gamma_\phi\,\omega$. 
For a displacement in the $(r,z)$ plane, the drag has components parallel and perpendicular to the membrane, 
$\gamma_\parallel = \gamma_o C_\parallel(s,R_b)$
and 
$\gamma_\perp = \gamma_o C_\perp(s,R_b)$,
respectively (Fig.\ref{fig_2}c). The correction terms $C_\perp, C_\parallel$  are defined in the SI.
As we focus on the movement of the bead along $\theta$, the corresponding drag is the projection
\begin{equation}
\gamma_\theta = L^2\sqrt{(\gamma_\perp^2 \cos^2\theta + \gamma_\parallel^2 \sin^2\theta)}.
\label{eq_gammatheo}
\end{equation}
We note that for each measurement, all the parameters discussed above and defined in Fig.\ref{fig_2}, can be quantified from the measured $x(t),y(t)$ position of the bead, after determination of $s_{\mbox{\tiny min}}$.

\begin{figure}
\centering
 \includegraphics[width=.8\linewidth]{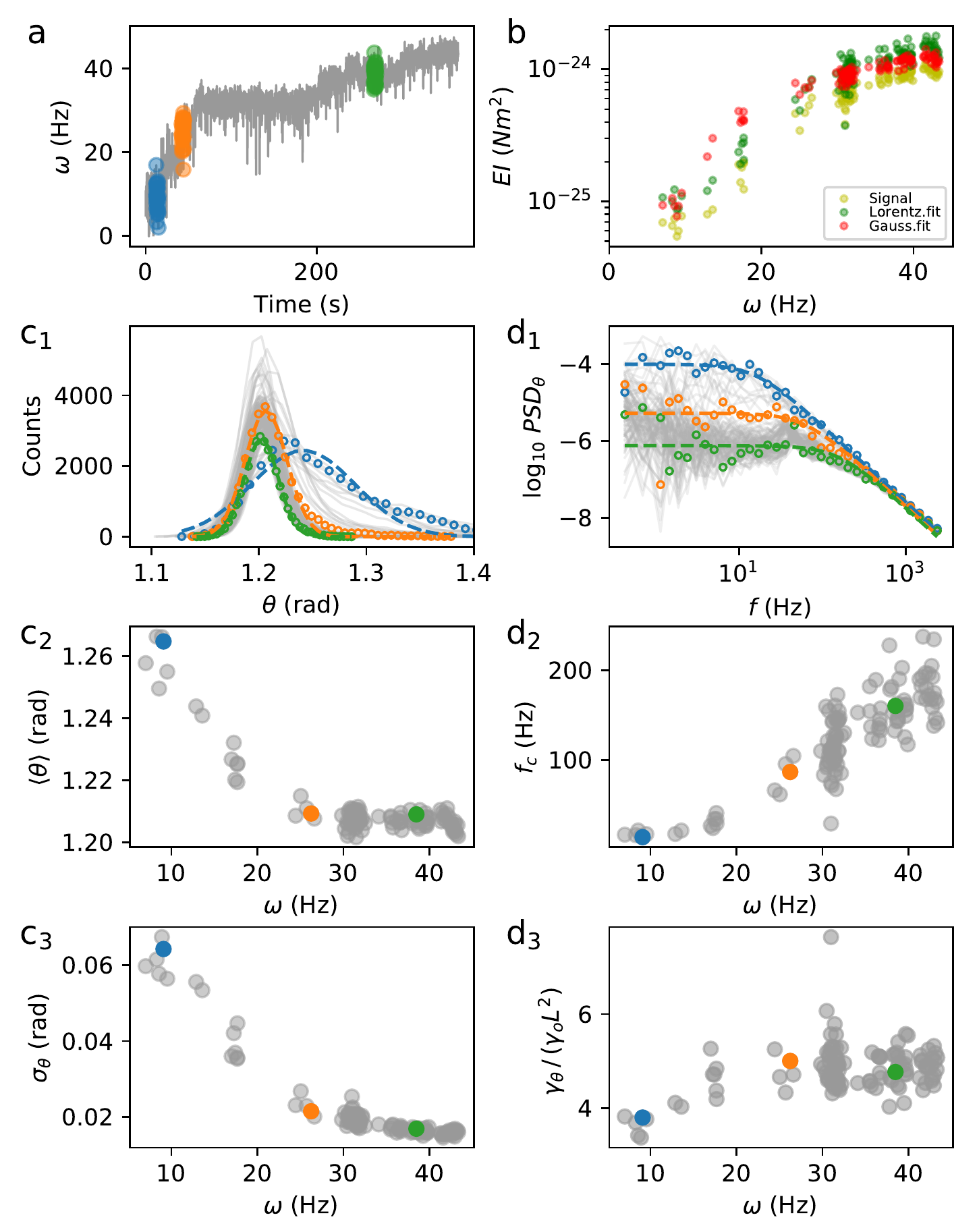}
\caption{Single motor fluctuation analysis. 
a) One resurrection trace of the speed $\omega$ ($R_b=500$ nm) is divided in time-windows of 4$s$ to allow local fluctuation analysis of the signal $\theta(t)$. Three windows are highlighted, and their colors are used for the corresponding points in all the panels.
b) Bending stiffness $EI$ as a function of speed $\omega$, calculated in each time-window following the analysis described in the text, using the raw signal $\theta_i(t)$, the Gaussian fit to its probability distribution, and the Lorentzian fit to its power spectral density.
c1) Probability density of all the $\theta_i$ along the trace (gray lines, while the points and their colors correspond to the windows shown in a), fit by a Gaussian function (dashed lines for the three windows highlighted).
c2) Mean value $\langle \theta_i \rangle$ as a function of the mean speed $\omega$ in the time-window.
c3) Standard deviation of $\theta_i$ as a function of the mean speed $\omega$ in the time-window.
d1) Power spectra $PSD_{\theta_i}(f)$ (gray lines) of all the time-windows along the trace, and Lorentzian fit (dashed lines) for the three highlighted windows.
d2) Corner frequency $f_c$ of the Lorentzian fit in d1 as a function of speed $\omega$.
d3) Drag coefficient $\gamma_{\theta}$ from the Lorentzian fit, normalized to the bulk value $\gamma_o L^2$.
}
\label{fig_3}
\end{figure}

\subsection*{Fluctuation analysis}
The radius $r(t)$ and the related angle $\theta(t)$ describe the motion of a point thermally fluctuating in a potential well in a rotating reference frame. By analyzing the fluctuations of the signals, the shape of the potential as a function of time can be dynamically characterized, yielding information on the elastic properties of the physical tether.
In order to characterize the fluctuations in $\theta$, we divide each experimental trace in time windows (Fig.\ref{fig_3}a), and in each window $i$ we calculate the mean and variance of the signal $\theta_i(t)$ from its probability distribution, which is well fit by a Gaussian function (Fig.\ref{fig_3}c1-c3), indicating a harmonic potential whose stiffness $\kappa_{\theta}$ can be quantified by the Equipartition theorem.
We then calculate the power spectral density of $\theta_i$, $PSD_{\theta_i}(f)$, a function of frequency $f$  (Fig.\ref{fig_3}d1-d3), which provides information on the stiffness and drag coefficient. 
Borrowing from the analysis commonly employed for objects in optical traps fluctuating in harmonic potentials \cite{neuman2004optical}, we fit the experimental $PSD_{\theta_i}(f)$ with the Lorentzian function $L(f) = (k_BT)/(\pi^2\gamma_{\theta}(f^2+f_c^2))$, where $k_BT$ is the thermal energy (Fig.\ref{fig_3}d1). The fit yields the drag $\gamma_{\theta}$ defined above (Fig.\ref{fig_3}d3) and the corner frequency $f_c$ (Fig.\ref{fig_3}d2), related to the potential stiffness by $\kappa_{\theta}=2\pi \gamma_{\theta} f_c$. 
The duration of the time-windows was chosen to sufficiently sample the plateau of $PSD_{\theta_i}(f)$ at low frequency, while the high sampling rate of the camera allows to sample frequencies higher than $f_c$ (Fig.\ref{fig_3}d1).
As the example in Fig.\ref{fig_3}c1-c3 shows, for an increasing motor speed $\omega$, the distribution of $\theta$ tends to become sharper, closer, in average, to the membrane (i.e. $\langle \theta \rangle$ decreases), and better approximated by a Gaussian.
At the same time, both the corner frequency $f_c$ and the drag $\gamma_{\theta}$ increase (Fig.\ref{fig_3}d2-d3). This reflects the bead being pushed towards the membrane for increasing $\omega$, where the drag becomes higher because of surface proximity, while the elastic tether becomes stiffer in bending.
In the SI, we use Langevin simulations to show that these observations, and in particular the increase in the measured corner frequency and stiffness, cannot be explained solely by the hydrodynamic increase in drag due to the proximity of the membrane. We also show that the centrifugal force cannot explain the observed bead displacement. Thus, our measurements point to an increasing bending stiffness of the hook.

\begin{figure}
\centering
\includegraphics[width=.8\linewidth]{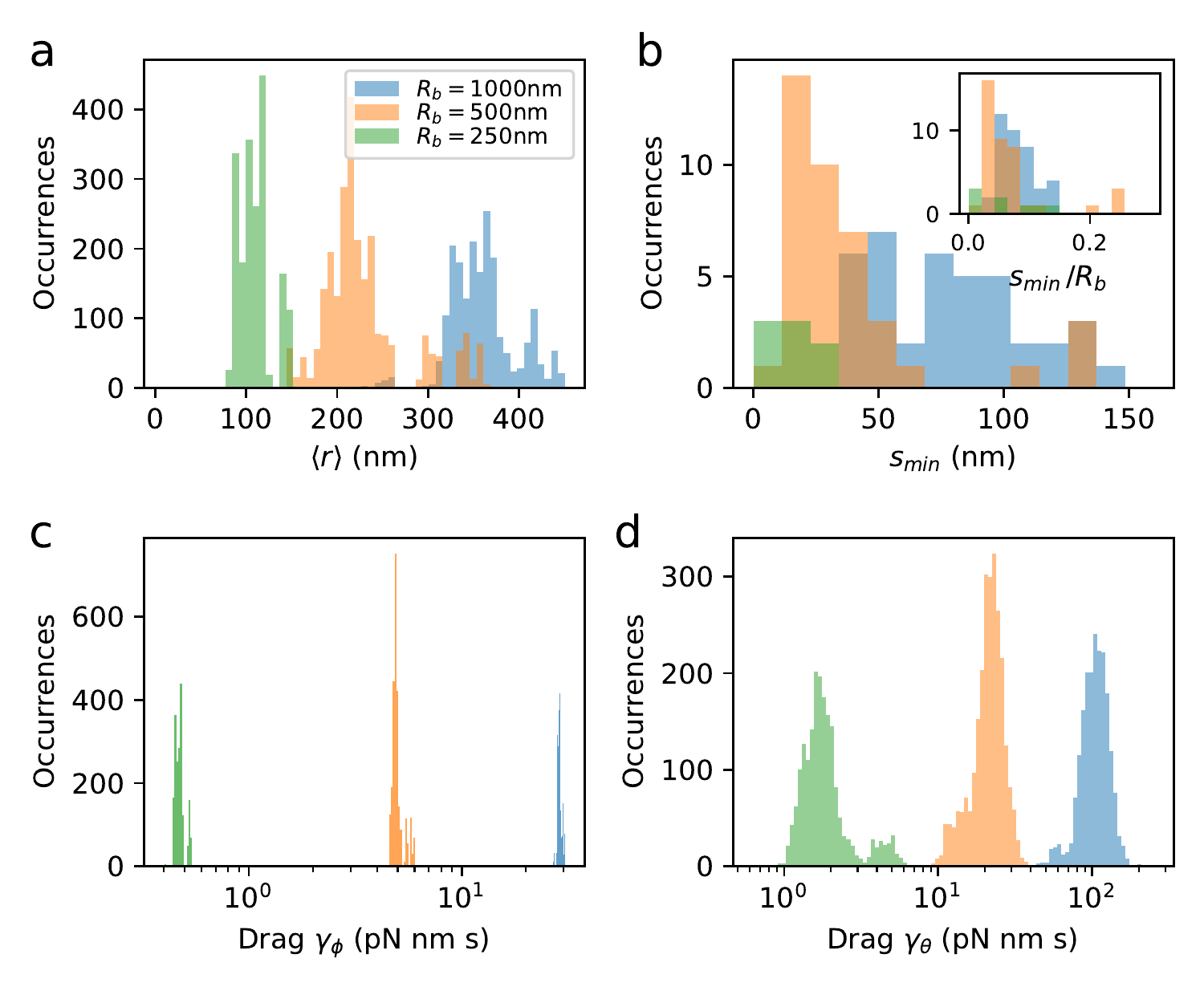}
\caption{Experimental distributions of parameters defined in Fig.\ref{fig_2}, obtained from the analysis of all the measured traces for the three loads considered. 
a) Histograms of the average radius $r_i$ in each time window of the $x,y$ bead trajectories. 
b) Histogram of the optimal value of the distance between the bead and the membrane  $s_{\mbox{\tiny min}}$, where one value is obtained from the analysis of each trace.
Inset: distributions of $s_{\mbox{\tiny min}}$ normalized by the value of the bead radius $R_b$.
c)-d) Distributions of the drag coefficients $\gamma_{\phi}$ and $\gamma_{\theta}$, obtained in each time window of all the traces. The color code, indicating the bead radius, is the same for all the panels.}
\label{fig_4}
\end{figure}

As mentioned above, the values obtained for $\theta$, $s$, $L$, $f_c$, $\gamma_{\phi}$, and $\gamma_{\theta}$ in each time-window depend on the choice of the distance $s_{\mbox{\tiny min}}$, which was left as a free parameter (Fig.\ref{fig_2}b). 
Not being directly measurable, we estimate $s_{\mbox{\tiny min}}$ by comparing the theoretical value of the drag $\gamma_{\theta}$ of eq.\ref{eq_gammatheo} (which depends on $s_{\mbox{\tiny min}}$) with the experimental value obtained from the Lorentzian fit of the spectrum (see SI).
The full procedure, described in detail in the SI, allows us to determine the values of all the quantities defined above from the experimental traces $x(t), y(t)$. In the SI, we also describe how the analysis is modified for the traces where we have also the trace $z(t)$, where the assumption of a constant $L$ (necessary only when $x,y$ are detected) is justified by the data. 

In Fig.\ref{fig_4} we plot the probability distributions for $\langle r \rangle$, $s_{\mbox{\tiny min}}$, and the two drag coefficients $\gamma_{\phi}$ and $\gamma_{\theta}$, resulting from the analysis performed on all the measured traces for three bead sizes. As the bead radius $R_b$ decreases, the average radius $\langle r \rangle$ of the trajectory decreases proportionately (Fig.\ref{fig_4}a and Fig.\ref{fig_1}b), as well as the value of the minimum distance $s_{\mbox{\tiny min}}$ (Fig.\ref{fig_4}b). When scaled by the bead radius, the three distributions of $s_{\mbox{\tiny min}}$ are similar (Fig.\ref{fig_4}b, inset), indicating that the bead minimum distance from the membrane is of the order of $0.1 R_b$.
Finally, the measured value of the drag $\gamma_\phi$ (Fig.\ref{fig_4}c), calculated using $s_{\mbox{\tiny min}}$, is used to calculate the motor torque ($\tau=\gamma_\phi \omega$, see fig.\ref{fig_5}).
The experimental values of the drag $\gamma_{\theta}$ in the plane $(r,z)$ (Fig.\ref{fig_4}d) result from the procedure used to determine $s_{\mbox{\tiny min}}$, and therefore are in agreement with the theoretical values obtained by  eq.\ref{eq_gammatheo}.

\subsection*{Hook bending stiffness}
The hook is the most flexible part of the tether which can bend and twist. While the angle $\theta$ does not correspond to the real bending angle of the hook, their variations are the same.
Therefore, the bending stiffness of the hook can now be determined from the fluctuations of $\theta_i(t)$ in each time window $i$ of the trace. Assuming a length of the hook $L_{\mbox{\footnotesize{hook}}}=60$ nm \cite{son2013bacteria}, the bending stiffness $EI$  (where $E$ is Young’s modulus and $I$ is the area moment of inertia) can be quantified using the Equipartition theorem by 
$EI = k_BT L_{\mbox{\footnotesize{hook}}} / {\sigma^2_{\theta_i}}$, where the variance $\sigma^2_{\theta_i}$ can be found either from the raw $\theta_i$ signal or from its Gaussian fit. 
Alternatively, the bending stiffness can be determined in each time window from the values of $\gamma_{\theta}$ and $f_c$ provided by the Lorentzian fit of the spectrum, by $EI = 2\pi\gamma_{\theta} f_c L_{\mbox{\footnotesize{hook}}}$. The three calculations (different but not fully independent) lead to similar results, as shown in Fig.\ref{fig_3}b for a single trace, where  $EI$ increases by one order of magnitude from $10^{-25}$ to $10^{-24}$ Nm$^2$, as the motor speed increases. Resolving with high resolution the speed changes due to stator incorporation helps us here to quantify $EI$ over a range of speed and torque values.

\begin{figure}
\centering
\includegraphics[width=.70\linewidth]{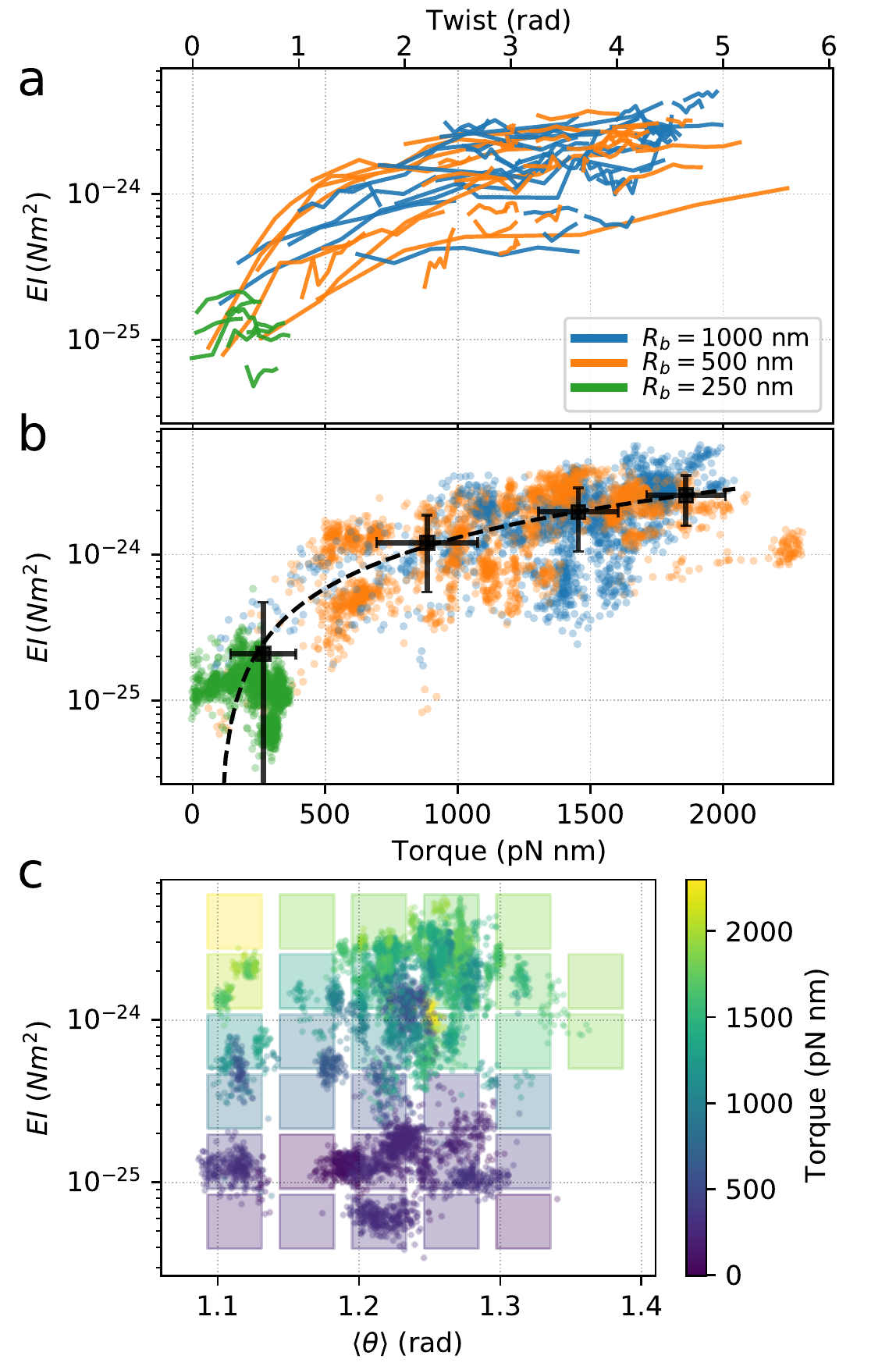}
\caption{Hook bending stiffness. 
a),b) Bending stiffness as a function of motor torque $\tau$ (bottom axis) and twist angle (top axis, calculated from a constant value of the torsional stiffness), for the three loads considered. In a) the lines indicate the trajectories  followed by individual motors during resurrection. In b) all the experimental points (one for each time-window) are shown, and are binned together in 4 points (black). The dashed line is a linear fit to the black points.
c) The experimental points are shown in the space ($EI$,$\langle\theta\rangle$,$\tau$), where $\langle\theta\rangle$ is a proxy for the bending angle, and the torque $\tau$ for the twist angle. The background colors indicate the average torque value in the corresponding region of the plane ($EI$,$\langle\theta\rangle$).}
\label{fig_5}
\end{figure}

At steady rotation, the torque produced by the motor is stored as twist of the hook. 
Such a torsional elastic link has a low-pass filtering effect on the measured position and speed of the bead, and is particularly problematic for the resolution of the fundamental step of the motor \cite{sowa2005direct, nord2018kinetic}.
In Fig.\ref{fig_5} we characterize the hook from all the traces acquired for the different loads as a function of motor torque. 
Using the published value of the hook torsional stiffness ($k_\phi = 400$ pN nm/rad \cite{block1989compliance, block1991compliance}), as a first approximation we can convert motor torque to twist, where the maximum torque of 2000 pNnm corresponds to a maximum twist of 280° (likely overestimated, as this is beyond the linear elastic region of $\sim$ 180°-270° reported in \cite{block1989compliance, block1991compliance}).
Fig.\ref{fig_5}a shows the individual trajectories of each hook in the plane ($\tau$ or twist, $EI$), where the torque produced in each time-window of the traces is calculated via the corresponding corrected drag $\gamma_\phi$. 
Despite the heterogeneity, it is clear that when the torque of the motor, and therefore the twist of the hook, increases due to a sufficiently high load ($R_b=500,1000$ nm, orange and blue curves of Fig.\ref{fig_5}a), the stiffness $EI$ of an individual hook can increase by a factor $10-20$.
Fig.\ref{fig_5}b shows in the same plane all the experimental points considered (one per time-window of each trace), used to build the trajectories in Fig.\ref{fig_5}a. We note that the trajectories of all the different loads globally converge at low torque in the region $EI\sim 0.5-1 \cdot 10^{-25}$ Nm$^2$, which should be the closest to the bending stiffness of the torsionally relaxed \textit{E.coli} hook. This range is compatible with the value of $3.6\pm0.4\cdot 10^{-26}$ Nm$^2$ found in \textit{V. alginolyticus} \cite{son2013bacteria}. 
A simple linear function (dashed line) fits reasonably well the binned average points (dark) of the plane. 
In Fig.\ref{fig_5}c we show that our analysis can further characterize the stiffness $EI$ both as a function of twist and bending, reported by torque $\tau$ and angle $\langle \theta \rangle$ (averaged in each time window), respectively. 
The experimental points populate the 3-dimensional space ($EI$,$\langle \theta \rangle$, $\tau$), where the background colors indicate the average torque in each region.
The stiffening observed in Fig\ref{fig_5}a-b, where all the bending angles are included, is resolved here for different values of bending $\langle \theta \rangle$. Moving vertically in the plot at constant $\langle \theta \rangle$, $EI$ increases with increasing twist and torque (reported by the color of the points).
On the other hand, moving horizontally, an increasing bending ($\langle \theta \rangle$) at  constant twist (color) is not accompanied by an appreciable change in stiffness $EI$. In other words, the bending stiffness increases with twist but not appreciably with bending angle, in the explored range.

The above results show that the hook stiffens when twisted by a motor rotating counter clock wise (CCW). Given the complex coiled structure of the hook \cite{shibata2019torque, kato2019structure}, we asked whether the hook would respond asymmetrically to twist applied in opposite directions. To compare the bending stiffness of a single hook twisted in both directions, we employed a mutant strain ($\Delta$CheRB) which  could switch the rotation direction of the motor, while maintaining CW and CCW speeds dwell times for sufficiently long  at different stator occupancy (see SI).
The results, shown in the SI, indicate that in a majority of cases the bending stiffness under CCW twist was higher than under CW twist. However, the heterogeneity of these results is large and further investigations are required to give a definitive answer and resolve a possible asymmetry.

\section*{Discussion}
Bead assays, in which a particle tethered to the flagellum of a rotating motor serves as a proxy for motor activity, have taught us much about the physics of the BFM over the past couple decades. Here, we show that the radial fluctuations of the tethered particle can dynamically report upon the mechanical properties of the flagellar hook.
Our high resolution data and novel fluctuation analysis reveal the hook to be a strain-stiffening polymer, showing a linear increase in hook bending stiffness, by more than one order of magnitude, as a function of the torque produced by the motor.

The hook, in order to function as a universal joint, is often described as a very soft element. 
To better visualize the rigidity of the polymer, the persistence length $L_p$ can be calculated from the bending stiffness $EI$ and the thermal energy $k_BT$, as $L_p=EI/k_BT$.
Our measurements show (in general agreement with those of \cite{son2013bacteria}) that the stiffness increases in the range $EI\sim 5\cdot 10^{-26} - 3\cdot 10^{-24} $ Nm$^2$, yielding a persistence length of about 10 $\mu$m in the hook's relaxed state, and up to several hundreds of microns in the torsionally loaded state. Considering other tubular biopolymers, this sets the hook between microtubules  ($L_p>1$ mm) and F-actin ($L_p\sim$ 10 $\mu$m) \cite{wen2011polymer}). 
It is worth noting that the long lever arm of the flagellum is relevant: for example, a 1 pN force applied along the flagellum 1 $\mu$m from the relaxed hook would produce a bending of 70 degrees.
Moreover, the area moment of inertia $I$ (calculated  for a hollow cylinder from the cross section as $I=\frac \pi 4 (R^4-r^4)$, where $R$ and $r$ are the external and internal radii, respectively) allows us to estimate the Young's modulus of the hook from the measured $EI$, which we find in the range $10^{7}-10^{9}$ Pa, depending on the choice of $R$ and $r$ (this simple result is for a homogeneous material, while the hook displays radial inhomogeneity \cite{flynn2004theoretical}). This is higher, by one to two orders of magnitude, than theoretical predictions \cite{flynn2004theoretical}, which are dependent on the experimental value, and its uncertainty, of the hook shear modulus $G$  \cite{block1991compliance}.

Over the range examined, for a fixed motor torque, our data show that the bending stiffness $EI$ is independent of the bending angle $\langle\theta\rangle$ (Fig.\ref{fig_5}c). Thus, for a fixed torsional stress, the hook behaves as a linear spring. Yet, with increasing twist of the hook (a consequence of increasing motor torque) the bending stiffness increases. What mechanism can explain this dynamic torsion-induced stiffening? 
One explanation may be a torsion-induced global restructuring of the hook, affecting its mechanical properties. The hook protein, FlgE, has three domains, and the 11 protofilaments of the hook form a short segment of a superhelix \cite{samatey2004, fuji2009, kato2019structure, shibata2019torque}; thus, each protofilament adopts a different length and its subunits a different repeat distance,  which varies periodically with motor rotation.  An atomic model of FlgE from \textit{Salmonella}, built from cryoEM single particle image analysis, has shown  11 distinct conformations of FlgE, with the subunits of a given protofilament having the same conformation. While there is little change in the conformation of each individual domain from one subunit to the next, a superposition of the 11 subunits shows a change in the relative domain orientations \cite{kato2019structure, shibata2019torque}. The inside bend of the hook is successively occupied by different protofilaments, and the compression and extension of each protofilament with each revolution arises due to dynamic changes in the relative orientations of the three FlgE domains about two hinge regions. This, in turn, begets dynamic changes in intersubunit interactions, which confer upon the hook its superhelical form and bending flexibility. Hook curvature is produced by close packing interactions between D2 domains along the 6-start helix on the inner side of the bend \cite{samatey2004, fuji2018}. The hook undergoes polymorphic transformations in response to changes in the temperature, salt concentration, or pH \cite{kato1984}, and it is thought that the superhelical curvature and twist depend on the direction of these D2-D2 interactions.  The D2 domain of subunit 0 also interacts with the D1 domain of subunit 11, and MD simulations have shown large axial sliding upon compression and extension \cite{shibata2019torque, samatey2004, kato2019structure}. 
An intrinsically disordered region that connects D0 and D1 governs inter-subunit interactions and has been shown to play a role in the stability of the hook structure \cite{barker2017intrinsically, hiraoka2017, matsunami2016complete}. It is therefore conceivable that, under the strain imposed by a global twist, changes in these interactions could give rise to decreased bending flexibility and increased bend, and this hypothesis could be explored by future MD simulations.

The fact that stiffening in bending occurs with increasing twist suggests that a coupling exists between these two classically decoupled directions. Experiments measuring the mechanical properties of other helical and supercoiled biopolymers such as actin filaments and DNA are well described by models which include a twist-bend coupling term \cite{enrique2010origin, nomidis2017twist}. However, we note that this formalism does not explain a change in stiffness (see SI). We also observe that for increasing torque the bead generally tends to be pushed towards the membrane. Twist-bend coupling produces an equilibrium bending angle that is directly modified by the twist in the structure (see SI), and therefore could be responsible for the approach of the bead to the surface.
However, such behavior of the bead could also be explained by the fact that, during rotation, the bead would tend to rotate around the axis of the tilted flagellar stub end, eventually being pushed against the membrane. 

Our results are in general agreement with the first observations of hook stiffening in the polar \textit{V. Alginolyticus} \cite{son2013bacteria}, where the measurements were performed on two conditions of the hook (relaxed and at physiological swimming-induced twist), and where the controlled buckling of the hook was shown to play a functional role for flicking, allowing this monotrichious bacterium to change swimming direction. The fact that a dynamic stiffening also occurs in multi-flagellated bacteria like \textit{E. coli}, may indicate that bundle formation and tumbling could benefit from the same mechanism as flicking, and that dynamic stiffening might be a common strategy in motile bacteria.
We expect that novel single-molecule force and torque manipulation essays will provide further insights into the mechanics of this striking biopolymer.

\section*{Methods}
\subsection*{Bacterial strains and growth}
The \textit{Escherchia coli} strain used was MT03 (parent strain: RP437, $\Delta pilA$, $\Delta cheY$, \textit{fliC::Tn10}), a non-switching strain with a `sticky' flagellar filament. For experiments which investigated $EI$ as a function of rotation direction, we used a strain in which we deleted \textit{cheR} and \textit{cheB} from MT02 (parent strain: RP437, $\Delta pilA$, \textit{fliC::Tn10}; see SI for details). Bacteria cultures were seeded from frozen aliquots (grown to saturation and stored in 25\% glycerol at  $-80^\circ$C) and grown in tryptone broth for 5 hours at 30$^\circ$C, shaking at 200 rpm, until an OD$_{600}$ of 0.5 - 0.8.  The flagellar filaments were sheared by passing the culture through a 21 G needle 50 times with a syringe. The culture was then washed and resuspended in motility buffer (MB, 10 mM potassium phosphate, 0.1 mM EDTA, 10 mM lactic acid, pH 7.0).
To vary the hydrodynamic load, affecting the speed of the BFM, we employ polystyrene beads (Sigma-Aldrich) of diameter 2000, 1000, and 500 nm .

\subsection*{Experimental measurements}

Custom microfluidic slides consisted of two coverslips (Menzel-Gl\"aser \#1.5) sealed by melted parafilm. The top coverslip had two holes for fluid exchange. Poly-L-lysine (Sigma-Aldrich) was introduced to the microfluidic slide, left to incubate for 5 min, then washed out with MB. Cells and then beads were sequentially introduced and allowed to sediment for 10 min, and the remnants were washed away with MB. Experiments were performed at 22$^\circ$C.
Rotating beads were imaged with a custom inverted in-line holographic microscope setup \cite{dulin2014efficient}. The sample was illuminated by a 660 nm laser diode (Onset Electro-Optics; HL6545MG) and imaged via a 100× 1.45 NA objective (Nikon) onto a CMOS camera (Optronics CL600x2/M) at 8.8 or 10 kHz.  The $x,y$ position of the bead (vectors parallel to the coverslip) were determined by cross correlation with a synthetic bead hologram. The z position of the bead (orthogonal to the coverslip) was determined by comparing the radial profile of the bead to profiles previously acquired at a known z via piezo controlled movement of the objective \cite{gosse2002}.
The bead trajectory is corrected to remove deterministic features and artifacts (see SI).
Resurrection experiments were performed by introducing carbonyl cyanide \textit{m}-chlorophenyl hydrazone (CCCP, Sigma-Aldrich) into the microfluidic slide for 5 min, then washing it out with MB. Data acquisition and bead tracking were performed with custom Labview scripts.

\subsection*{Data analysis}
Each experimental trace was divided in time windows of 1, 3, and 4 s for beads of radius $R_b=250,500$, and 1000 nm, respectively. In each window, the angle $\theta(t)$ (see Fig.\ref{fig_2}) was calculated from the ($x(t),y(t)$)  or ($x(t),y(t),z(t)$) position of the bead.  The angular fluctuations of the hook were fit with a Lorentzian function to extract the bending stiffness of the hook, $EI$. Full details of the analysis workflow are given in SI. All analysis was performed with custom Python scripts.

\subsection*{Acknowledgements} 
We thank Richard Berry, Nils-Ole Walliser, Luca Costa, and Marcelo Nollmann for fruitful discussions. The bacterial strain used in this work were gifts from the labs of Judy Armitage and Richard Berry. ABB and TP were supported by the ANR FlagMotor project grant ANR-18-CE30-0008 of the French \textit{Agence Nationale de la Recherche}. The CBS is a member of the France-BioImaging (FBI) and the French Infrastructure for Integrated Structural Biology (FRISBI), 2 national infrastructures supported by the French National Research Agency (ANR-10-INBS-04-01 and ANR-10-INBS-05, respectively).



\newpage

\begin{center}
{\Large \textbf{Supplementary Information}}
\vspace{10pt}

{\Large Dynamic stiffening of the flagellar hook revealed by radial fluctuation analysis in bead assays}

\vspace{10pt}

Ashley L. Nord,
Ana\"{i}s Biquet-Bisquert, 
Manouk Abkarian,
Théo Pigaglio,\\
Farida Seduk,
Axel Magalon,
Francesco Pedaci
\end{center}

\vspace{50pt}

\tableofcontents

\clearpage

\section{Supplementary figures}

\begin{figure}
\centering
\includegraphics[width=.95\linewidth]{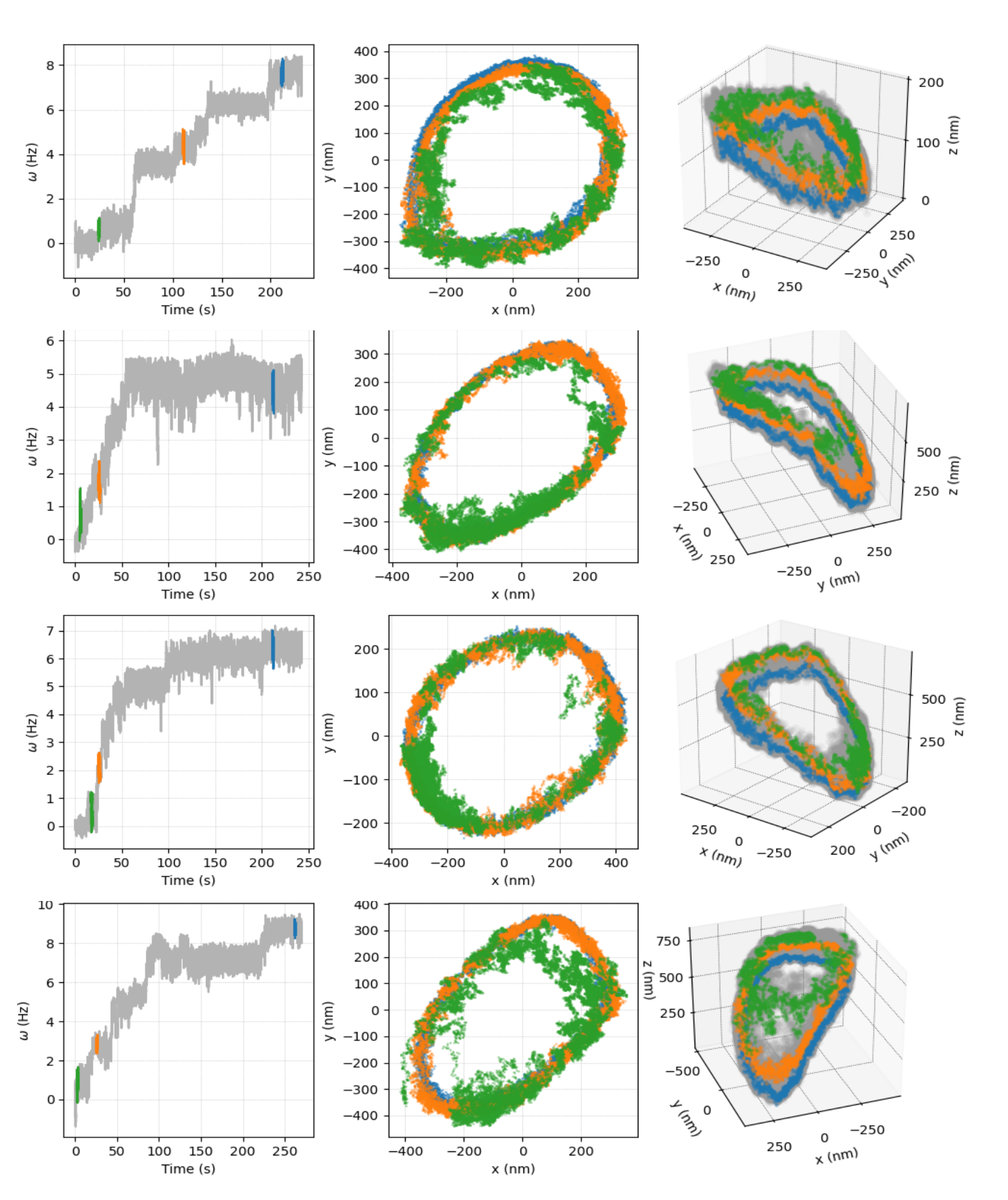}
\caption{Examples of 3D trajectories of beads ($R_b=1000$ nm). Each line corresponds to a different bead. The left column shows the angular speed $\omega(t)$, where the resurrection of the motor is evident. Three particular time-windows are highlighted in green, orange and blue. The middle column shows the $xy$ projection of the trajectory, where the colors correspond to the highlighted time-windows. At early times (green) the rotation of the bead is more erratic. The right column shows the trajectory in 3 dimensions.}
\label{fig_SI3dtrajex}
\end{figure}

\begin{figure}[ht]
\centering
\includegraphics[width=.8\linewidth]{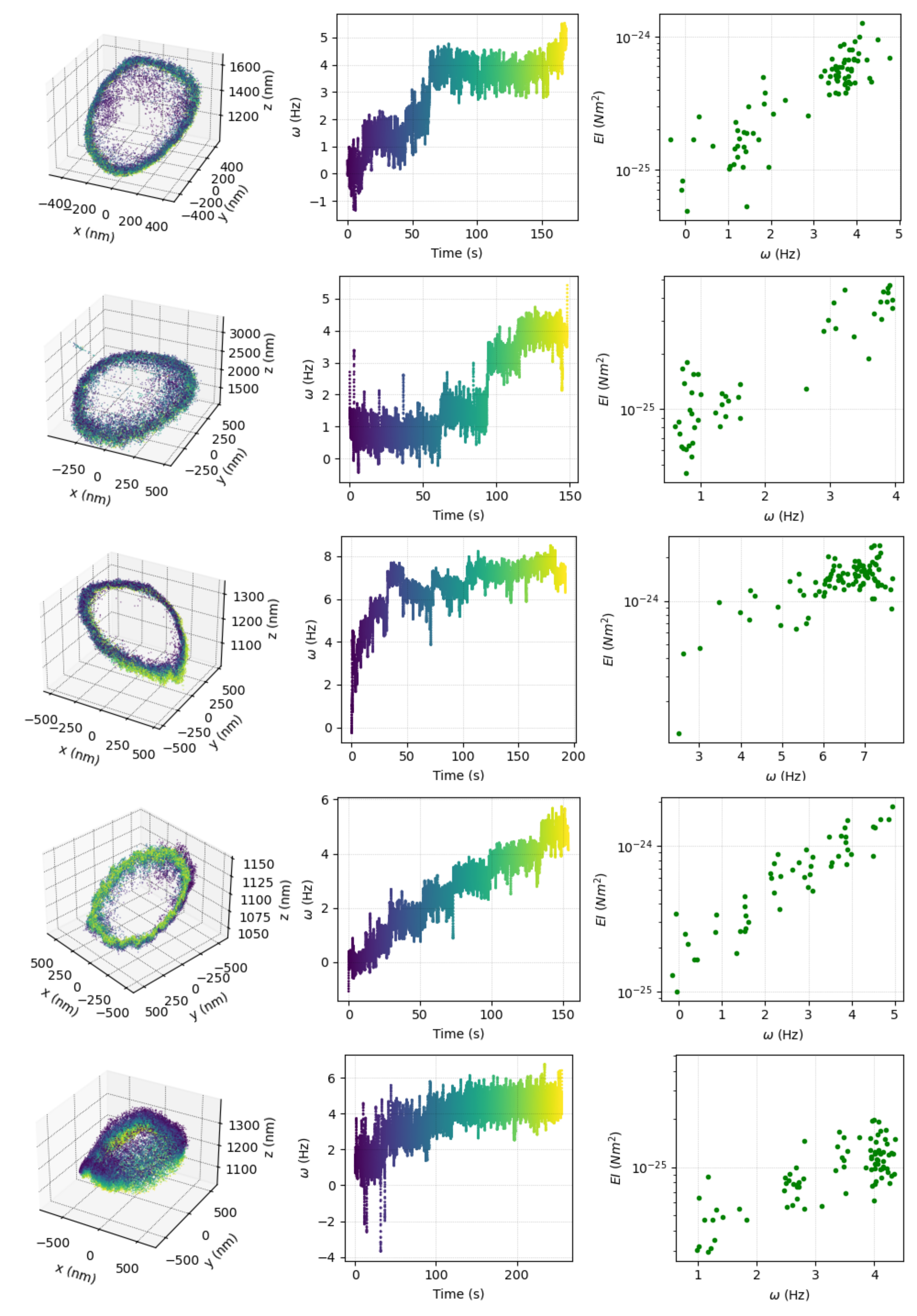}
\caption{Examples of the result of the analysis leading to the value of the bending stiffness $EI$ taking into account the full 3D trajectory of the bead. Each line corresponds to a different motor bound to a $R_b=1000$ nm bead. In each row, the left panel shows the recorded 3D trajectory of the bead center. The middle panel shows the angular speed $\omega(t)$ during resurrection. In the left and middle panels the color from dark to yellow indicates time. The right panel shows the resulting bending stiffness $EI(\omega)$ calculated considering the 3D trajectory, and defining $\theta$ and $L$ from $x,y,z$ (For more details, see the ``case $xyz$'' in the analysis workflow of sec.\ref{SIworkflow}). }
\label{fig_SI3DEI}
\end{figure}

\begin{figure}[ht]
\centering
\includegraphics[width=.99\linewidth]{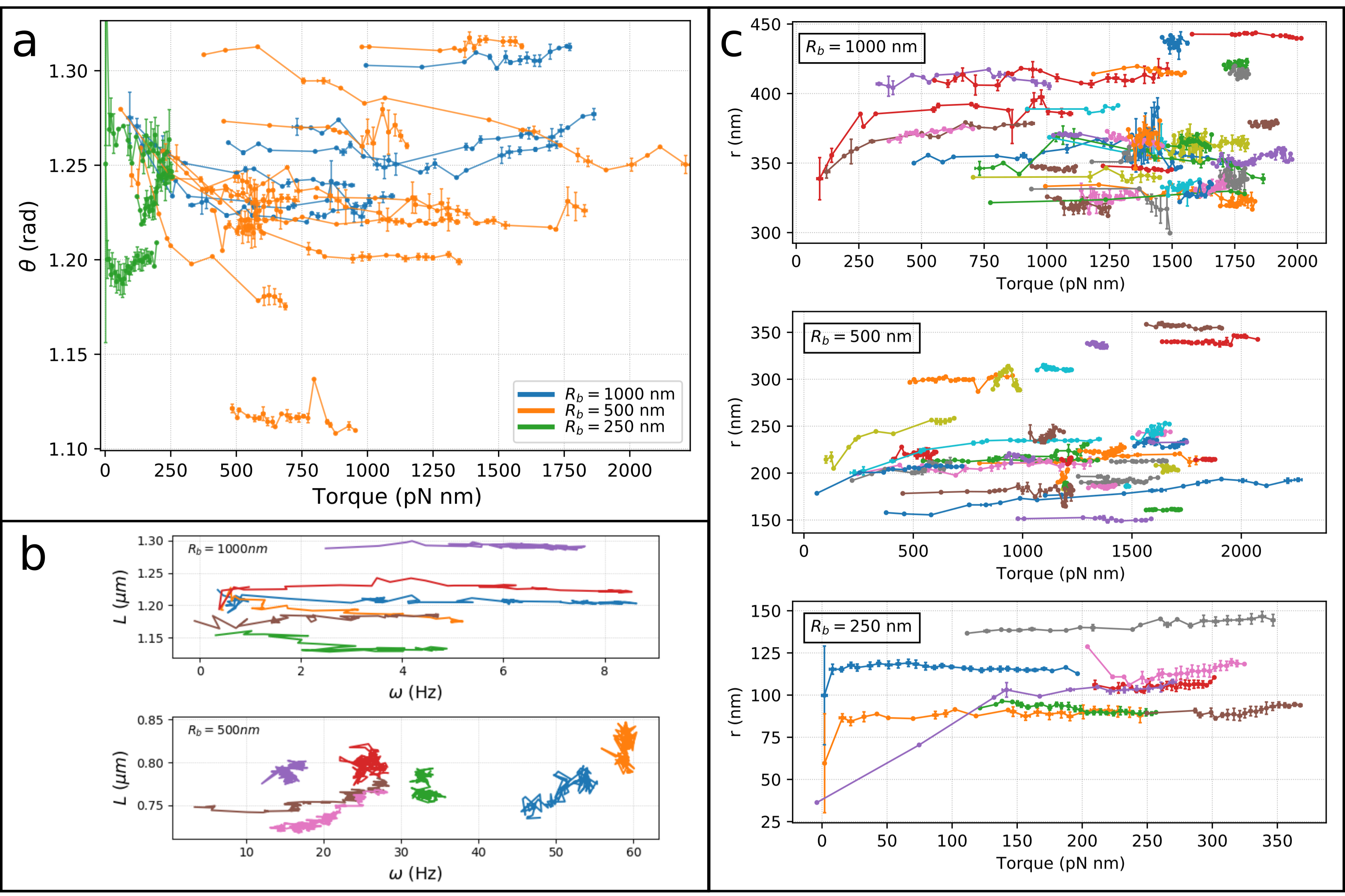}
\caption{Geometrical analysis of all the traces recorded (see Fig.\ref{SIfig_2} for the geometrical definitions). Based on these observations we formulate the simplifying hypothesis that the bead rotates on hemispherical dome, approaching the membrane as the rotation speed increases.
a) The angle $\theta$, averaged in each time-window of each trace, is shown as a function of the measured motor torque. The three colors indicate the three bead sizes employed, as indicated. Generally, and especially starting from low torque, as the motor accelerates during resurrection, the bead moves down towards the membrane, reflected by a decrease in $\theta$. 
b) Values of $L$ (distance between bead center and hook origin on the membrane, averaged on short 0.5-1 s time-windows along the traces) extracted from $x,y,z$ bead trajectories, as a function of angular speed, for beads with $R_b=1000,500$ nm. For a given bead in a 3D trajectory, L remains reasonably constant.
c) For the same data shown in a) we show the radius $r$ of the $x,y$ trajectory, averaged in each time-window, as a function of motor torque. The movement of the bead towards the membrane is reflected here by an increase in $r$ for a particular trajectory. The three loads are split in the three panels, as indicated.}
\label{fig_SI_thetaVstq}
\end{figure}

\clearpage
\section{Drag coefficients}
\label{SIsec:DragCoeff}

\begin{figure}
\centering
\includegraphics[width=.7\linewidth]{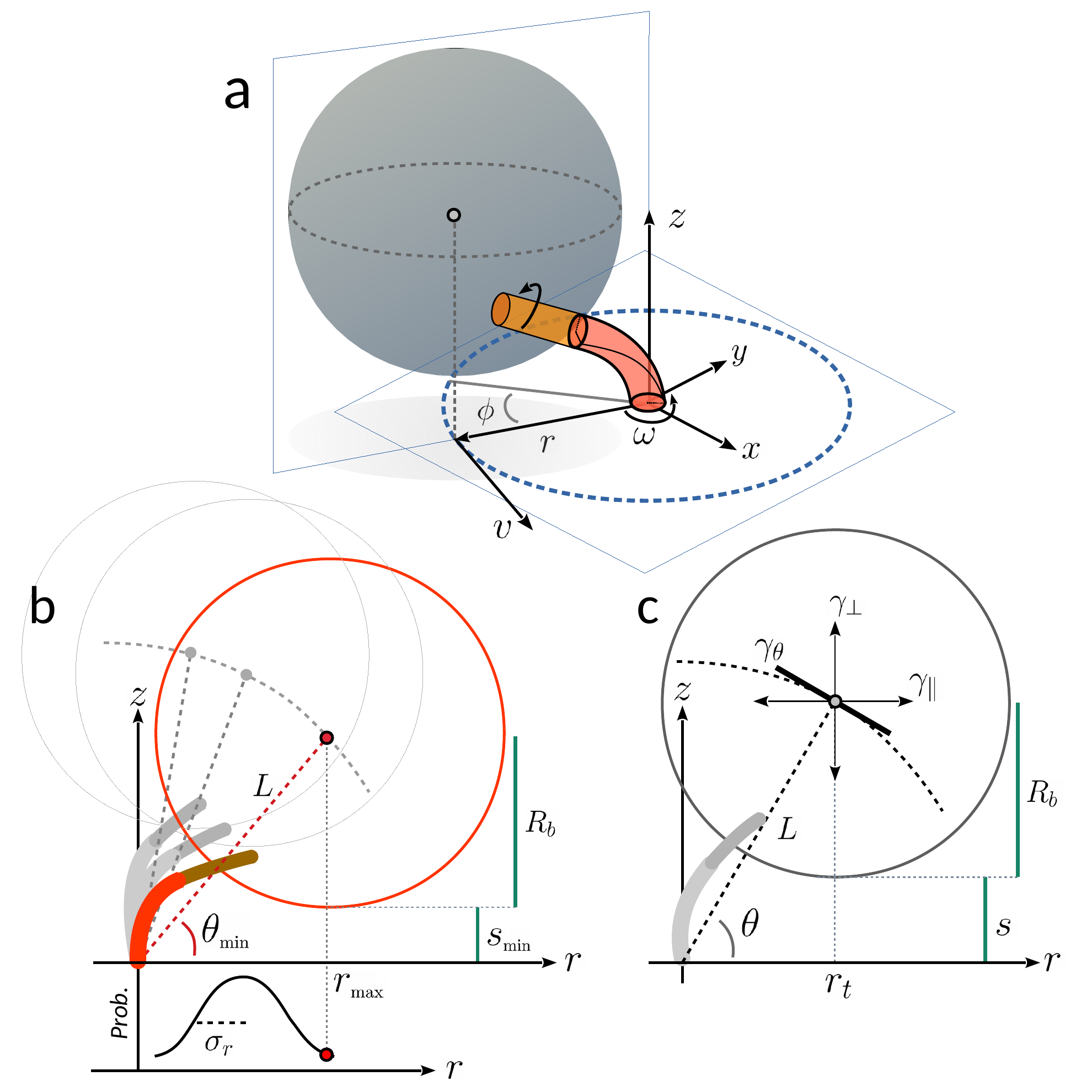}
\caption{Microscopic geometrical model. }
\label{SIfig_2}
\end{figure}

\subsection{Plane (r,z)}
We aim at writing the expression of the bead angular drag coefficient $\gamma_{\theta}$ in the direction tangent to the arc trajectory of radius $L$, in the plane $(r,z)$, described by the angle $\theta$ (Fig.\ref{SIfig_2}).
We consider the movement of the center of the bead along the arc, where the linear tangential speed is $v_{tg}=\dot\theta L$, and where the force acting on the bead is $F_{tg}=v_{tg}\gamma_{tg}$, with $\gamma_{tg}$ the linear drag coefficient in the tangential direction. The associated torque can then be written as $\tau_{\theta}=\gamma_{\theta} \dot\theta = F_{tg}L = v_{tg}\gamma_{tg} L = \gamma_{tg} L^2 \dot\theta$. The linear drag $\gamma_{tg}$ is composed by the parallel and perpendicular components, as $\gamma_{tg}^2 = \gamma_{\parallel}^2 \sin^2(\theta) + \gamma_{\perp}^2 \cos^2(\theta)$. Therefore, the angular drag $\gamma_{\theta}$ can be written as
\begin{equation}
    \gamma_{\theta} = \gamma_{\theta}(\theta, s, R_b) = L^2 \sqrt{\gamma_{\parallel}^2 \sin^2(\theta) + \gamma_{\perp}^2 \cos^2(\theta)}. \label{eqSI_gamma_theta}
\end{equation}
The components $\gamma_{\parallel}$ and $\gamma_{\perp}$ can be written following the treatment developed by Faxen or Brenner, given below. In Fig.~\ref{fig_SI_FaxenBrenner_theta} we show the value of these components as a function of the gap $s$ between the wall and the bead (of radius $R_b=500$ nm). In our analysis we calculate $\gamma_{\theta}$ from both theories (Faxen and Brenner), and for every $\theta$ (or $s$) we take the maximum of the two.

\subsubsection{Faxen expressions}
Following the theory by Faxen \cite{leach2009comparison, neuman2004optical}, the drag components can be written as,
\begin{eqnarray}
\gamma_{\parallel,F} & = & \frac{\gamma_o}{1 - \frac{9}{16}\frac{R_b}{d} + \frac 1 8 (\frac{R_b}{d})^3} \label{eq:SIgamma_perp_paral1_faxen} \\
\gamma_{\perp,F} & = & \frac{\gamma_o}{1 - \frac 9 8 (\frac{R_b}{d}) + \frac 1 2 (\frac{R_b}{d})^3},
\label{eq:SIgamma_perp_paral2_faxen}
\end{eqnarray}
where $\gamma_o = 6\pi\eta R_b$ is the drag of a spherical particle in the bulk (far from surfaces), $\eta$ is the medium viscosity, $R_b$ is the radius of the particle, and $d=s+R_b$ is the distance between the bead center and the wall.

\begin{figure}
\centering
\includegraphics[width=.6\linewidth]{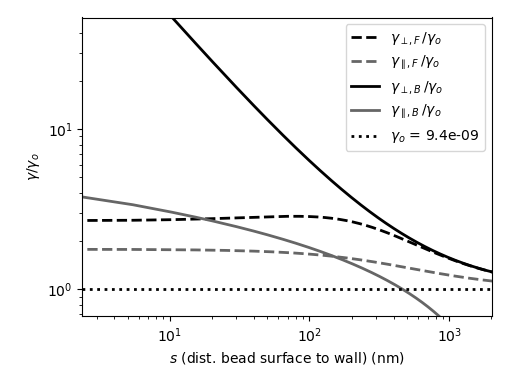}
\caption{Comparison between the expressions of the drag components of a bead with radius $R_b=500$ nm from Faxen (eqs.\ref{eq:SIgamma_perp_paral1_faxen}, \ref{eq:SIgamma_perp_paral2_faxen}) and Brenner theory (eq. \ref{eq:SIgamma_perp_paral1_brenner},  \ref{eq:SIgamma_perp_paral2_brenner}) as indicated in the plot legend. The bulk drag is  $\gamma_o=6\pi\eta R_b$ and its value is indicated in the legend in Ns/m.}
\label{fig_SI_FaxenBrenner_theta}
\end{figure}

\subsubsection{Brenner expressions}
Following Brenner theory \cite{goldman1967slow, brenner1961slow}, the drag components are, 
\begin{eqnarray}
\gamma_{\perp,B} & = &\gamma_o \, C_\perp(s,R_b) 
\label{eq:SIgamma_perp_paral1_brenner}\\
\gamma_{\parallel,B} & = & \gamma_o \, C_\parallel(s,R_b),
\label{eq:SIgamma_perp_paral2_brenner}
\end{eqnarray}
where $R_b$ is the radius of the spherical particle. $C_{\perp}, C_{\parallel}$ are correction factors for the bulk drag $\gamma_o$, functions of $R_b$ and of the gap $s$ between the bead surface and the membrane. Their expressions read,
\begin{eqnarray}
C_\parallel(s,R_b) &=& \frac{8}{15}\ln \left( \frac{s}{R_b} - 0.9588 \right) \\
C_\perp(s,R_b) &=& \frac{4}{3}\sinh(\alpha) \sum_n \frac{n(n+1)\;C_n}{(2n-1)(2n+3)},
\end{eqnarray}
where, in the expression of $C_\perp(s,R_b)$, we use $n=1,...,20$ and 
\begin{eqnarray}
C_n & = & \left[ \frac{2\sinh((2n+1)\alpha) + (2n+1)\sinh(2\alpha)}{4\sinh^2((n+\frac{1}{2})\alpha) - (2n+1)^2\sinh^2(\alpha)} -1   \right] \\
\alpha & = & \ln \left( \frac{s+R_b}{R_b} + 
\sqrt{\left(\frac{s+R_b}{R_b} \right)^2-1} \, \right). \end{eqnarray}

\begin{figure}
\centering
\includegraphics[width=.99\linewidth]{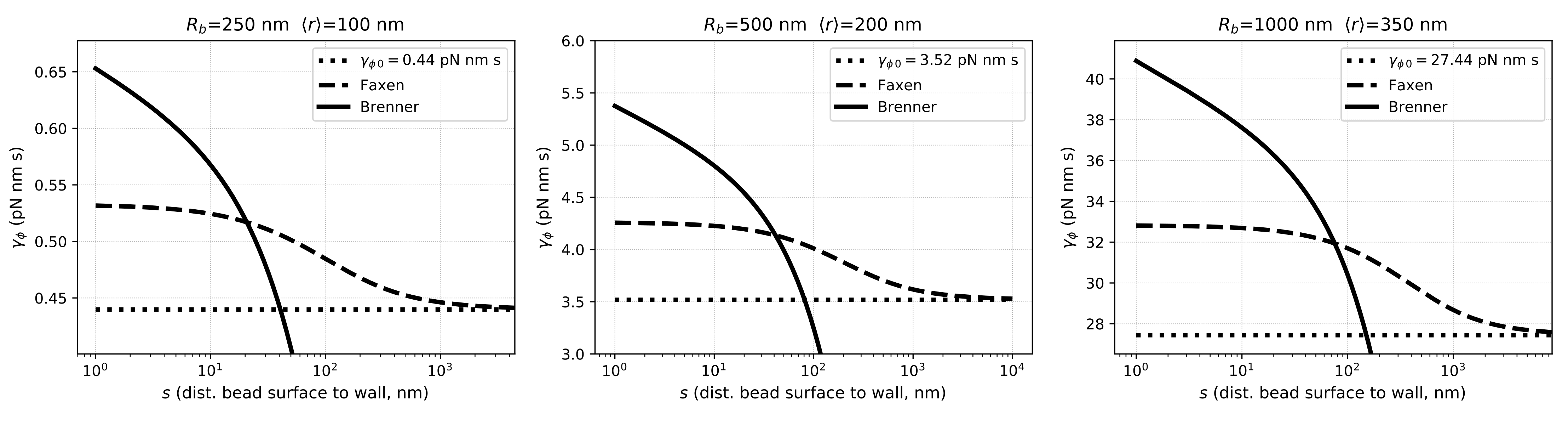}
\caption{Comparison between the expressions of $\gamma_{\phi,F}$ (Faxen, eq.\ref{SI_eq_gamma_phi_faxen}) and $\gamma_{\phi,B}$ (Brenner, eq.\ref{SI_eq_gamma_phi_brenner}) as a function of the gap $s$ between the bead and the wall, for a bead of radius $R_b$ translating and rotating on a circular trajectory (parallel to the plane $x,y$) of radius $\langle r \rangle$ (with the same face pointing the center), as indicated in the title of each panel. The bulk drag is defined as $\gamma_{\phi\,0} = 8\pi\eta R_b^3 + 6\pi\eta R_b\langle r \rangle^2$}
\label{fig_SI_FaxenBrenner}
\end{figure}

\subsection{Plane parallel to (x,y)}
For the rotation described by the angle $\phi$ (Fig.~\ref{SIfig_2}a), the drag $\gamma_{\phi}$ includes one component for the rotation of the bead around its axis (proportional to $8\pi\eta R_b^3$), and one component for its rotation along a circle of radius $\langle r \rangle$ (proportional to $6\pi\eta R_b \langle r \rangle^2$).
Two formalisms provide an analytical expression for the drag corrected due to proximity with a rigid wall.
The Faxen expression for the corrected drag coefficient is valid for $s \ge R_b$, and can be written as \cite{leach2009comparison}
\begin{equation}
\gamma_{\phi,F}  = \frac{8\pi\eta R_b^3}{1 - \frac 1 8 (\frac {R_b}{R_b+s})^3}  + \frac{6\pi\eta R_b \langle r \rangle^2}{1 - \frac{9}{ 16} (\frac{R_b}{R_b+s}) + \frac 1 8 (\frac{R_b}{R_b+s})^3} \hspace{15pt}.
\label{SI_eq_gamma_phi_faxen}
\end{equation}
The expression due to Brenner is valid for $s\ll R_b$ and reads
\begin{eqnarray}
\gamma_{\phi,B} & = & 8\pi\eta R_b^3 \; \left( 1.20205 - 3 \left( \frac{\pi^2}{6} - 1 \right)\frac s R_b \right)    + 6\pi\eta R_b \langle r \rangle^2  \; \left(\frac{8}{15} \log(\frac{s}{R_b}) - 0.9588 \right)
\label{SI_eq_gamma_phi_brenner}
\end{eqnarray}
We note that the Faxen expression $\gamma_{\phi,F}$ has the advantage to remain finite for every value of $s$, while $\gamma_{\phi,B}$ diverges outside its range of validity (Fig.~\ref{fig_SI_FaxenBrenner}). For $s/R_b \rightarrow 0$, the Faxen expression $\gamma_{\phi,F}$ (outside its range) remains lower than the more accurate Brenner expression $\gamma_{\phi,B}$ by $\sim 20 \%$ for the beads we employ.
Therefore, despite the widespread use of $\gamma_{\phi,F}$ (e.g. in the optical trapping and BFM literature), one should be careful not to employ it for distances much smaller than the bead radius, where $\gamma_{\phi,B}$ should be preferred, in order to not under estimate the motor torque ($\tau=\gamma_{\phi} \omega$). In our analysis, we can extract the distance $s$ in each time window of the traces, and for a given $s$ we use the drag,
\begin{equation}
\gamma_{\phi}(s) = \max \left( \gamma_{\phi,F}(s), \gamma_{\phi,B}(s) \right). \label{SIeqmaxFaxBren}
\end{equation}

\section{Analysis workflow}
\label{SIworkflow}
We describe here in detail the workflow followed to extract all the parameters from the experimental traces. The data consist in the tracked position of the center of the bead $x(t),y(t)$ (labeled `case $xy$' in the analysis below). In a smaller number of cases we also have the $z$ position of the bead (labeled `case $xyz$'). In this case, the analysis can rely directly on $z(t)$. Overall, the goal of the analysis described below is to extract the angle $\theta(t)$ in small non overlapping time-windows $\theta_i(t)$ along the trace, which reflect the changes in time of the locally averaged bending of the hook. Non overlapping windows form a set of independent measurements, and are beneficial to decrease the computational time.
The fluctuations of $\theta_i(t)$, via its probability distribution and spectrum, together with our geometrical assumptions, provide a measurement of the hook bending stiffness $EI$, and its variation with motor speed, motor torque and hook twist change.

For each trace we run the following analysis, described below as a function in pseudo code termed \verb+radial_analysis()+, which accepts as inputs the traces $x(t),y(t)$ (optionally $z(t)$), and the offset distance  $s_{\mbox{\footnotesize{min}}}$. 
As the $x,y,z$ bead positions are measured relatively to an arbitrary origin, the distance between the bead and the membrane is unknown. One goal of the analysis is to estimate this distance by quantifying the offset $s_{\mbox{\footnotesize{min}}}$.
Once we determine $s_{\mbox{\footnotesize{min}}}$, all the geometrical variables can be determined from $x,y,(z)$. In particular: 
\begin{itemize}
    \item in each time-window $i$, $\theta_i(t)$ and the fit of its spectrum provides the experimental value of the drag $\gamma_{\theta,i}$
    \item the distance $d_i$ between the bead center and the membrane can be fixed allowing the calculation of the theoretical drag $\gamma_{\theta,th,i}$ in the time window. 
\end{itemize}

For an arbitrary choice of $s_{\mbox{\footnotesize{min}}}$, the values of the experimental $\{\gamma_{\theta,i}\}$ and theoretical $\{\gamma_{\theta,th,i}\}$  drag, along the time windows of one trace, are different. For example, if $s_{\mbox{\footnotesize{min}}}$ is assumed to be too large, the theoretical $\{\gamma_{\theta,th,i}\}$ will contain only the contribution of the bulk drag, while the measured $\{\gamma_{\theta,i}\}$ will be larger, due to the proximity of the bead to the membrane. 
For a given input $s_{\mbox{\footnotesize{min}}}$, the function \verb+radial_analysis(+$s_{\mbox{\footnotesize{min}}}$\verb+)+ calculates the values of $\{\gamma_{\theta,i}\}$ and $\{\gamma_{\theta,th,i}\}$ along the input trace, and the mean square error (MSE) between them. A subsequent automatic procedure, calling \verb+radial_analysis(+$s_{\mbox{\footnotesize{min}}}$\verb+)+ multiple times, varies the parameter $s_{\mbox{\footnotesize{min}}}$ in order to minimize the MSE. 
The value of $s_{\mbox{\footnotesize{min}}}$ that minimizes the MSE is finally retained.

\vspace{30pt}

Function \verb+radial_analysis+($x,y(,z)$ [arrays], $s_{\mbox{\footnotesize{min}}}$ [floating point]):
\begin{enumerate}
    \item (optional) case $xyz$: remove outlier points in $z$ (which can arise when the tracker fails to converge on one frame)
    \item remove drift in $x,y(,z)$ using a trace simultaneously recorded of a surface-immobilized bead located in the same field of view
    \item Scale the entire $x,y$ trajectory into a circle, fitting it to an ellipse and scaling the minor axis to become equal to the major axis 
    \item case $xyz$: 
    \begin{enumerate}
        \item multiply $z$ by the refraction index correction factor (0.85) \cite{hell1993aberrations}
        \item (optional) correct the 1-turn periodic modulation of $z$ (arising from a non-circular trajectory, see SI sec.\ref{SIsec:correcction})
        \item modify the value of $z$ by shifting it vertically, $z \rightarrow z - min(z) + R_b + s_{\mbox{\footnotesize{min}}}$, so $z$ indicates the distance between the bead center and the cell membrane. The values of $z$ depend now on the choice of $s_{\mbox{\footnotesize{min}}}$
        \item define $\theta = \arctan(z/ \sqrt{x^2 + y^2})$ \label{item:theta_xyz}
    \end{enumerate}

    \item given window size ($0.5-4$ s depending on the size of the attached particle), set the time windows from which the windowed arrays $x_i,y_i,(z_i,\theta_i)$ are defined
    \item case $xy$: 
    \begin{enumerate}
        \item on each window $i$, center the trajectory: $x_i \rightarrow x_i - \langle x_i \rangle$, and $y_i \rightarrow y_i - \langle y_i \rangle$
        \item on each window, find the values of the radius $r_i=\sqrt{x_i^2+y_i^2}$.
        \item find the value of $r_{\mbox{\footnotesize{max}}}$ (one value for the entire trace)
        \item find the value of $L=\sqrt{(R_b + s_{\mbox{\footnotesize{min}}})^2 + r_{\mbox{\footnotesize{max}}}^2}$ (one value for the entire trace) \label{item_Lxy}
    \end{enumerate}
    \item Main loop. On each time window $i$:
    \begin{enumerate}
        \item center the trajectory: $x_i \rightarrow x_i - \langle x_i \rangle$, and $y_i \rightarrow y_i - \langle y_i \rangle$
        \item calculate the values of the following arrays
        \begin{itemize}
            \item $\phi_i = \arctan({y_i/x_i})$, the tangential angle
            \item $\omega_i = d\phi_i/dt$, the angular speed of the bead
            \item $r_i = \sqrt{x_i^2+y_i^2}$, the radius of the trajectory
        \end{itemize}
        \item correct the 1-turn periodic modulation of $r_i$ (SI sec.\ref{SIsec:correcction})
        \item find the values $\theta_i(t)$ of the angle $\theta$ in the current time-window $i$
        \begin{itemize}
            \item case $xy$: $\theta_i = \arccos(r_i/L)$, where $L$ is defined in \ref{item_Lxy}
            \item case $xyz$: $\theta_i$ windowed from $\theta$ defined in \ref{item:theta_xyz}
        \end{itemize}
        \item calculate $PSD_{\theta_i}(f)$, the single-sided power spectral density of $\theta_i(t)$, function of frequency $f$
        
        \item fit the experimental $PSD_{\theta_i}(f)$ with the theoretical expression $PSD(f) = \frac{k_BT}{\pi^2\gamma_{\theta}(f^2+f_c^2)}$ (we use the python function \verb+scipy.optimize.differential_evolution+), to obtain in each window $i$ the corner frequency $f_{c,i}$ and the experimental drag $\gamma_{\theta,i}$ \label{item_lorenfit}
        
        \item find the probability distribution of $\theta_i$, and fit it with a Gaussian function \label{item_gaussiandistrib}
        \item in the current time-window $i$, define $L_i$ (the local value of $L$) and $d_i$ (the local distance between the bead center and wall) as:
        \begin{itemize}
            \item case $xy$: $L_i = L$ (the global value defined in \ref{item_Lxy}), and $d_i = \sqrt{L^2 - \mbox{med}(r_i) ^2}$, where med() indicates the median
            \item case $xyz$: $L_i = \sqrt{d_i^2 + \mbox{med}(r_i)^2}$, and $d_i = \mbox{med}(z_i)$, where med() indicates the median
        \end{itemize}
        \item using SI eq.\ref{eqSI_gamma_theta}, calculate the theoretical value of the drag $\gamma_{\theta_i, \mbox{\footnotesize{th}}}(d_i)$ for the given $r_i, \theta_i,  L_i, s_{\mbox{\footnotesize{min}}}$ \label{item_gammatheo}
        
    \end{enumerate}
    
    \item \label{item:EI} Calculate in each time-window the bending stiffness of the hook (of length $L_{\mbox{\footnotesize{hook}}}=60$ nm) from the equipartition theorem, as:
    \begin{enumerate}
        \item $EI_{sig} = \frac{k_BT}{\sigma^2_{\theta_i}}L_{\mbox{\footnotesize{hook}}}$, where the variance $\sigma^2_{\theta_i}$ is calculated directly from the signal $\theta_i$
        
        \item $EI_{gaus} = \frac{k_BT}{\sigma^2_{\theta_i}}L_{\mbox{\footnotesize{hook}}}$, where $\sigma^2_{\theta_i}$ is variance of the Gaussian fit to the distribution of $\theta_i$, found in \ref{item_gaussiandistrib} 
        
        \item $EI_{lor} = 2\pi\gamma_{\theta}f_c L_{\mbox{\footnotesize{hook}}}$ where the drag $\gamma_{\theta}$ and the corner frequency $f_c$ are obtained from Lorentzian fit of $PSD_{\theta_i}(f)$ found in \ref{item_lorenfit}
        \end{enumerate}
        These three methods to estimate of $EI$ are not fully independent, and the difference between them is  used as an internal consistency check. Only the value of $EI_{lor}$ is kept in the following. This step \ref{item:EI} is relevant only for when the input $s_{\mbox{\footnotesize{min}}}$ is the optimal value.
    
    \item Calculate and return the value of  $\mbox{MSE}(\gamma_{\theta}, \gamma_{\theta, \mbox{\footnotesize{th}}})$, the Mean Square Error between the experimental and theoretical drag (found in \ref{item_lorenfit} and \ref{item_gammatheo}, respectively), a function of the choice of $s_{min}$
    
\end{enumerate}

As mentioned above, for each trace, the optimal value for the offset $s_{\mbox{\footnotesize{min}}}$ is found by automatically minimizing (we use the python function \verb+scipy.optimize.minimize+) the value of $\mbox{MSE}(\{\gamma_{\theta,i}\}, \{\gamma_{\theta, \mbox{\footnotesize{th},i}}\})$, returned by the function \verb+radial_analysis+($x,y(,z), s_{\mbox{\footnotesize{min}}}$), by varying the input value $s_{\mbox{\footnotesize{min}}}$ alone.
For each trace, after the optimal $s_{\mbox{\footnotesize{min}}}$ is determined, the value of $s_i$ (the gap between the bead and the membrane) can be found in each time window $i$ by $s_i = L_i^2-r_i^2 - R_b$. This is susequently used to calculate the tangential drag $\gamma_{\phi}(s)$ (SI eq.\ref{SIeqmaxFaxBren}) and the motor torque $\tau=\gamma_{\phi}\, \omega$.

\section{Corrections of the bead trajectory} \label{SIsec:correcction}

\begin{figure}
\centering
\includegraphics[width=.99\linewidth]{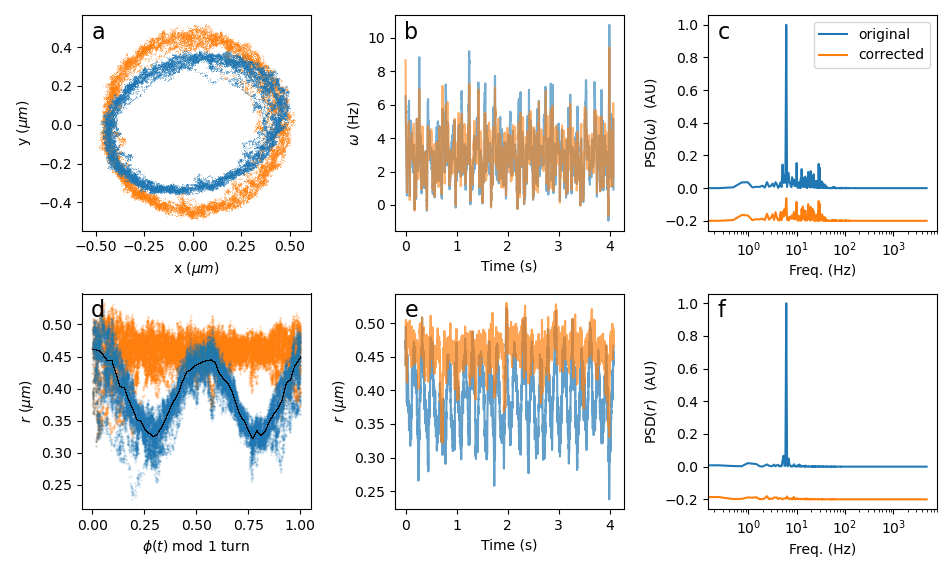}
\caption{Example of corrections performed on the experimental traces. In all panels, the original data are in blue, and the corrected data are in orange. In a-c the correction consists in fitting an ellipse on the trajectory and scale it to a circle. The spurious periodic modulation in the speed trace due to the ellipticity of the trajectory is in this way removed. 
In e-f the correction consists in removing directly from the signal (here the radius $r(t)$) the 1-turn periodic modulation.
a) Portion of the $x,y$ trajectory of a $R_b=1000$ nm bead.
b) Angular speed $\omega(t)$ of the bead.
c) Normalized spectrum of the speed $\omega(t)$, where the periodic modulation present in the original data (blue) is suppressed by the scaling of the trajectory (orange, vertically shifted for clarity). 
d) The original signal $r(t)=\sqrt{x^2+y^2}$ (blue) is strongly modulated every turn due to the ellipticity of the trajectory. By subtraction of a high degree polynomial fit, or equivalently of an interpolated signal, (black line) the modulation is mitigated (orange).
e) Time trace $r(t)$ with and without the correction shown in d.
f) Normalized spectrum of the original (blue) and corrected radius $r(t)$ (orange, vertically shifted for clarity).
}
\label{SIfig_corrections}
\end{figure}

In bead assays, the measured trajectory of the bead is rarely perfectly circular. It is very common to observe an elliptical $x,y$ trajectory, and this is usually explained by assuming the real 3D trajectory to be circular, but lying on plane tilted with respect to the image plane, as can occur when the motor is on the side of the cell. The projection onto the image plane would then be observed as an ellipse. Observing bead trajectories in three dimensions, we have observed that the real 3D trajectory can also be not perfectly circular as assumed. This is probably the result of the interaction of the bead with the local topography of the cell. 

Whenever the trajectory (either in 2D or 3D) differs from a circle, and if this perturbation is always present (as when it is due to the cell topography), the signals of interest obtained from the trajectory (e.g $z(t)$, $\omega(t)$, $r(t)$) acquire a modulation that is periodic with respect to the position of the bead along the trajectory. Due to this periodicity, these signals can be corrected, as we show in one example in Fig.\ref{SIfig_corrections}. 

In  Fig.\ref{SIfig_corrections}a we show a bead that displays an elliptical $x,y$ trajectory. 
We can perform two kinds of corrections on such traces: 
\begin{enumerate}
    \item (Fig.\ref{SIfig_corrections}a-c) We fit an ellipse to the $x,y$ trajectory, and, using the ratio of the major to minor axis, we transform the $x,y$ points into a circular trajectory. This alleviates the periodic modulation that otherwise affects the traces. In Fig.\ref{SIfig_corrections}a-c we show the effects of the correction on the speed trace $\omega(t)$ and its spectrum. In the time window shown, the angular speed $\omega(t)$ is strongly modulated at the frequency of rotation, as indicated by the peak of its spectrum. The modulation and the peak disappear after the correction.
    
    \item We correct directly the signal affected by the periodic modulation (we use $r(t)$ in Fig.\ref{SIfig_corrections}d-f) by plotting it as a function of $mod(\phi(t)/2\pi, 1)$, effectively wrapping the signal onto itself every turn, Fig.\ref{SIfig_corrections}d. The 1-turn periodic modulation can be high pass filtered by fitting and subtraction of a high order polynomial or of an interpolated signal. In Fig.\ref{SIfig_corrections}d-f, we show the effect of this procedure on the signal $r(t)$ and its spectrum.
\end{enumerate}

\section{Simulating a particle in a harmonic potential in presence of a drag gradient}

\begin{figure}
\centering
\includegraphics[width=1\linewidth]{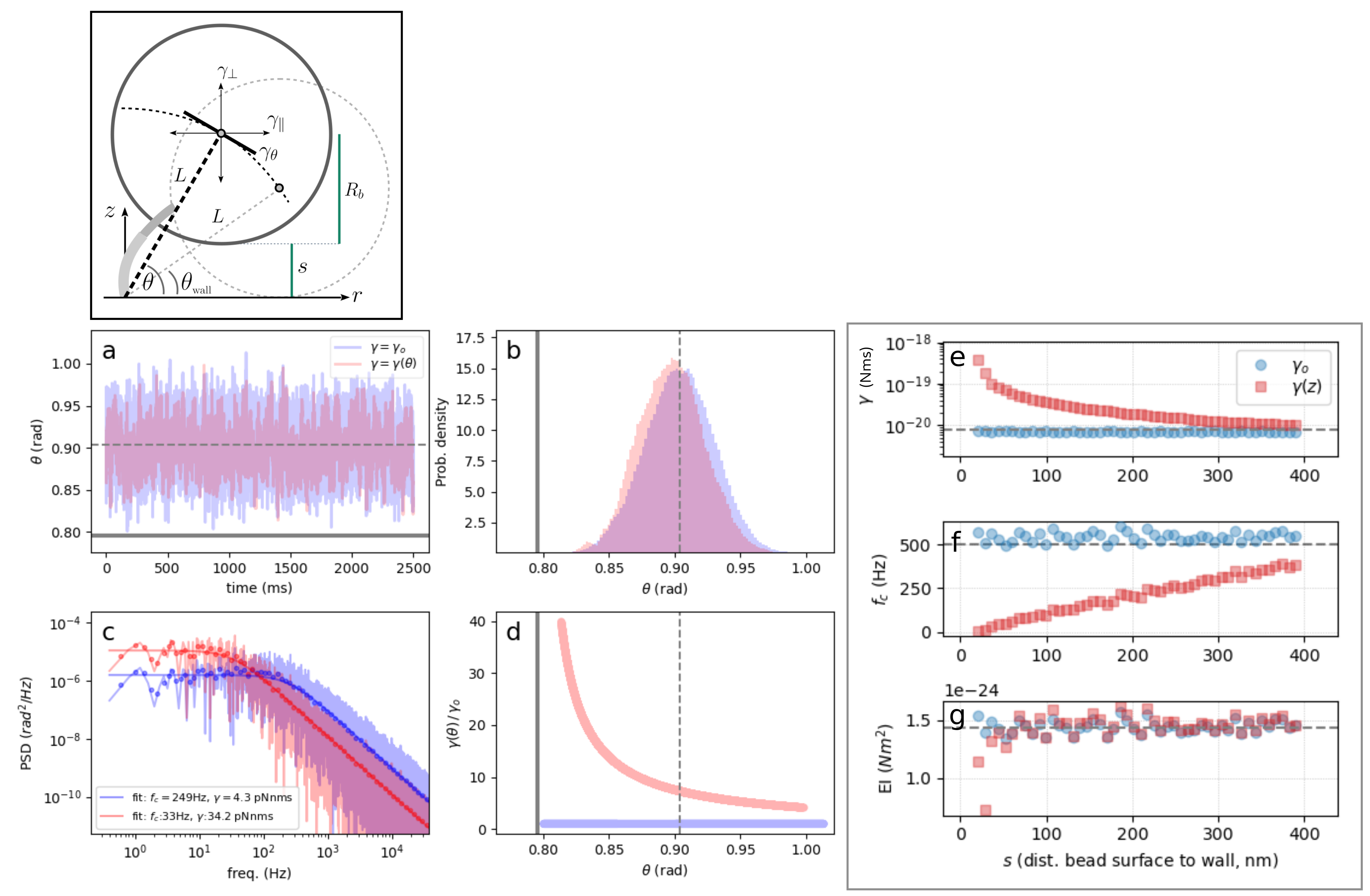}
\caption{Simulations of a particle in a harmonic potential with tailored drag dependency on position. 
\textbf{Top panel:} schematic geometry simulated, where the parameters have been chosen to be compatible with our measurements. A bead ($R_b=500$ nm) can move on an arc of circle of radius $L=700$ nm from the motor origin. 
The thermal fluctuations are simulated by a Langevin equation describing the evolution of the angle $\theta(t)$, which is assumed to be held at an equilibrium position by a harmonic potential centered at $\theta_o=0.9$ rad ($s=50$ nm, $r(\theta_o)=430$ nm). 
The bead collides with an absorbing wall at $\theta=\theta_{\tiny\mbox{wall}}$.
The harmonic potential has a stiffness $\kappa_{\theta}=2\pi f_c \gamma_{\theta}$ defined by the corner frequency $f_c=200$ Hz and a drag $\gamma_{\theta}$ which we set to either i) the constant bulk value $\gamma_o = 6\pi\eta R_b L^2=4.6$ pN nm s (blue curves and points) or ii) to the value of the function $\gamma_{\theta}(\theta)$ corrected by the presence of the wall by Brenner and Faxen theory (red curves and points). The noise term is delta-correlated.
\textbf{a-d)} Simulation and analysis. 
a) Simulated traces of $\theta(t)$ for the two choices of the drag $\gamma_{\theta}$. In the panels a,b,d the dashed line indicates the equilibrium angle $\theta_o$, and the thick line indicates the position of the wall $\theta_{\tiny\mbox{wall}}$. b) Probability distributions of the visited angle. c) Power spectral densities of $\theta$. A Lorentzian fit (line) is performed on the equally log-spaced points shown. The parameters obtained from the fit are shown in the legend. In the case of $\gamma_{\theta}=\gamma_o$ (blue), they compare well with the input parameters of the simulation. The effect of the change in drag as a function of $\theta$ is reflected by the shift of the red spectrum with respect to the blue. d) Profiles of the drag $\gamma_{\theta}(\theta)/\gamma_o$ used in the simulation. The blue line is fixed at the value of 1, while the red increases in direction of the wall.
\textbf{e,f,g)} Analysis results of simulations at varying distance $s$. The three panels show the  values returned by the fit of the PSD of $\theta$ for the drag $\gamma$ (panel e), the corner frequency ($f_c$, panel f), and the hook stiffness $EI=2\pi f_c \gamma L_{\tiny\mbox{hook}}$ (panel g), in simulations (like those shown in panel a) where the distance $s$ from the wall was changed, changing the equilibrium angle $\theta_o$.
}
\label{SIfig_OTsimul}
\end{figure}

A particle close to a rigid wall experiences an increasing drag as the distance to the wall decreases, as described by Faxen's and Brenner's equations (eq.\ref{SI_eq_gamma_phi_brenner},\ref{SI_eq_gamma_phi_faxen}). 
Here, we explore by simulations the effect of this gradient on the diffusion of a particle trapped in a harmonic potential. The potential in our case is provided by the hook considered as an angular spring, providing a restoring force directed towards the equilibrium position. A similar potential could be provided by an optical trap.
The analysis of the experimental data shows that, as motor speed and torque increase, the mean bead position moves towards the surface, the corner frequency $f_c$ of the Lorentzian fit increases, the measured drag $\gamma_{\theta}$ increases, and the stiffness of the potential increases (Fig.3 of the main text).
Here, we ask whether the increase in drag, induced by the wall proximity, can \textit{alone} explain these observations, therefore excluding the mechanism of hook stiffening. 

To answer this question, we have run Langevin simulations that reproduce our geometrical assumptions (we note that the choice of the algorithm is not trivial in the presence of viscosity gradients \cite{de2013translocation}). A bead or radius $R_b$ can move along an arc of radius $L$. The bead position is described by the angle $\theta(t)$, and is trapped in a harmonic potential centered at an equilibrium angle $\theta_o$. We can tailor the drag profile of the bead as a function of its position, and vary the equilibrium position in order to explore the behavior of limit cases. 
In Fig.\ref{SIfig_OTsimul}a-d, we allow the bead to fluctuate in the harmonic potential either i) in the absence of a wall, considering the constant bulk drag $\gamma_o$ (in blue in all the panels), and ii) in the presence of the wall ($s=50$ nm, red curves and points), where the drag dependency on $\theta$ is given by combining the components $\gamma_{\perp}$ and $\gamma_{\parallel}$ both from Brenner and Faxen formulas (at a given $\theta$, the drag $\gamma_{\theta}(\theta)$ is taken as the maximum between the expressions of Brenner and Faxen, see SI sec.\ref{SIsec:DragCoeff}. The resulting drag used in the simulation as a function of the angle $\theta$ for the two cases is shown in Fig.\ref{SIfig_OTsimul}d. 

We then treat the simulated $\theta$ displacement as an experimental trace. Its probability distribution, at the chosen distance $s=50$ nm from the wall, is affected (0.01 rad) with respect to the simulation in the absence of the wall, as the particle spends more time in the region of higher effective viscosity. This occurs also in the experiment, but we note that the experimentally measured drag is only 4-5 times higher than bulk, while in the simulation a similar shift of the probability distribution is achieved with a drag 10-30 times higher than bulk.
As in the experiment, in Fig.\ref{SIfig_OTsimul}c, we fit the PSD of the simulated $\theta(t)$ to a Lorentzian, obtaining a corner frequency $f_c$ and an effective drag $\gamma$. 
In absence of the wall (blue spectrum), the fit returns the input parameters as expected, within the error.
In proximity to the wall (red), despite the $\theta$-dependency of the drag, the spectrum (red in Fig.\ref{SIfig_OTsimul}) maintains overall a Lorentzian shape. With respect to the case of constant drag, the spectrum is now shifted, giving raise to both a reduced corner frequency $f_c$, and an increased effective drag $\gamma$ (see the legend of panel c). 
Due to the fact that the angular stiffness ($2\pi\gamma f_c L^2$) is proportional to both $f_c$ and $\gamma$, this results in a stiffness that does not change significantly in presence of the wall.
Moreover, while the increase of the fit drag goes in the direction of the experimental observation (although higher than in the experiment), a concomitant decrease of the corner frequency, and the resultant unaffected stiffness are not in agreement with our  experimental observations.
In panels e-g), we show the result of the PSD fit analysis performed while changing in the simulations the gap $s$ between the bead and the wall. In presence of the wall and for a decreasing $s$, the fit drag increases (red points in panel e), the fit corner frequency decreases (panel f), and the stiffness remains at the same value as in absence of the wall. This is in disagreement with the experimental observations.

In conclusion, these simulations and their analysis show that the hydrodynamic effects due to a rigid wall, while they can account for qualitative features like the shift of the distribution towards higher viscosity and an increased drag obtained from the PSD, fail to explain the increase both in corner frequency and stiffness observed in the experiment. Therefore, the dynamic stiffening of the hook remains a valid mechanism to explain our data.

\section{Bend-twist coupling and persistence length}
In the classical continuum description of a rod, bending and twisting are independent.
Following the treatment and the definitions used for DNA \cite{marko1994bending, nomidis2019twist}, a coupling between the two can be inserted writing the the elastic free energy as
\begin{equation}
    F = \frac{1}{2} k_BT \int_0^L [A(\Omega_1^2+\Omega_2^2) + C\Omega_3^2 + 2G\Omega_2\Omega_3] \, ds
\end{equation}
Here, an orthonormal frame of three vectors $\{e_1,e_2,e_3\}$ is used to characterize each point of the rod, where $e_3$ is tangent to the curve. Three corresponding rotation vectors $\{\Omega_1,\Omega_2,\Omega_3\}$ (dimensions [rad/m]) connect adjacent local frames ${e_i}$, and describe any deformation of the relaxed configuration. Variations in $e_1$ and $e_2$ indicate bending along the two orthogonal directions, while variations in $e_3$ indicate twist. The parameter $s$ is the arc-length of the curve, and $L$ is the total length. The lengths $A$, $C$, and $G$ denote the persistence lengths for bending, twist, and bend-twist coupling, respectively. $k_BT$ is the thermal energy. Assuming the simplest case of constant quantities to resolve the integral, taking $\Omega_1=0$ and defining the twist and bending angle respectively as $\theta_T=\Omega_3 L$ and $\theta_B=\Omega_2 L$, the energy can be written as 
\begin{equation}
    F = \frac{k_BT}{2L} (A \theta_B^2 + C \theta_T^2 + 2G \theta_B\theta_T)
\end{equation}
Twist and bend angle are confined in a parabolic potential well, with coupling. The restoring twist and bend torque can be written as
\begin{eqnarray}
\tau_{\,T} & = & - \frac{\partial F}{\partial \theta_T} = - \frac{k_BT}{L}(C\theta_T + G\theta_B)\\
\tau_{\,B} & = & - \frac{\partial F}{\partial \theta_B}  = - \frac{k_BT}{L}(A\theta_B + G\theta_T)
\end{eqnarray}
The presence of the coupling introduced in this manner adds an offset to the restoring torque. This is a term that, being not dependent on the variable, does not influence the stiffness (equal to $\partial \tau_i/\partial \theta_i$). Therefore, such coupling can shift the equilibrium angles, but cannot explain a change in stiffness.

In absence of coupling or for relaxed twist ($G\theta_T=0$), one retrieves the linear relationship between restoring torque and angle, which for bending reads $\tau_B=\frac{k_BT A}{L}\theta_B = \frac{EI_o}{L}\theta_B$, where $EI_o=k_BT A$ is the bending stiffness of the relaxed hook ($\theta_T=0$). As noted in the main text, our measurements are not performed on a perfectly twist-relaxed hook, but for low torque (low load and low speed, $\tau<100$ pNnm) our data ($EI_{o}= 1.2 \pm 0.4 \cdot 10^{-25}$ Nm$^2$) are compatible with the value found in torsionally relaxed hooks of \textit{Vibrio}, $EI_o=3.6 \pm 0.4 \cdot 10^{-26}$ Nm$^2$ \cite{son2013bacteria}. We note that the persistence length associated to this value of the bending stiffness is $A=EI_o/(k_BT)\sim 8\, \mu $m. As noted in \cite{son2013bacteria}, an $EI_o$ two orders of magnitude lower has been measured by electron microscopy \cite{sen2004elasticity}, yielding a persistence length of the order of the length of the hook. Our measurements and those described in \cite{son2013bacteria}, both based on the fluctuations of the hook in its native environment (and thus without the possible perturbations due to imaging in electron microscopy), indicate that the hook is probably not as soft as often pictured. However, we note that even with a persistence length of 8 $\mu$m, much longer than its dimensions, a stiffness $EI_o=3.6 \pm 0.4 \cdot 10^{-26}$ Nm$^2$ allows thermal fluctuations of 10-20 degrees in \textit{V.alginolyticus} \cite{son2013bacteria}. In Table \ref{table_EI} we summarize the existing measurements of the hook bending stiffness.

\begin{table}[t]
\small
\begin{center}
\begin{tabular}{l l l l}
 $EI$ (Nm$^2$) & Bacterial strain & Note & Ref. \\ 
 \hline
$1.2 \pm 0.4 \cdot 10^{-25} \rightarrow 27\pm 9 \cdot 10^{-25}$ & \textit{Escherichia coli} & Relaxed$\rightarrow$Loaded  & This study \\ 
$3.6 \pm 0.4 \cdot 10^{-26}$ & \textit{Vibrio alginolyticus} & Relaxed & \cite{son2013bacteria} \\  
$2.2 \pm 0.4 \cdot 10^{-25}$ & \textit{Vibrio alginolyticus} & Loaded  & \cite{son2013bacteria} \\  
$1.6 \cdot 10^{-28}$         & \textit{Escherichia coli}     &         & \cite{sen2004elasticity} \\
$3.0 \cdot 10^{-28}$         & \textit{Salmonella typhimurium} &       & \cite{sen2004elasticity} \\
$4.0 \cdot 10^{-28}$         & \textit{Vibrio cholerae}      &         & \cite{sen2004elasticity} \\
$4.8 \cdot 10^{-28}$         & \textit{Vibrio parahaemolyticus} &      & \cite{sen2004elasticity} \\
$5\cdot 10^{-28} - 5\cdot 10^{-27}$ & \textit{Salmonella typhimurium} & Theoretical & \cite{flynn2004theoretical} \\
\hline
\end{tabular}
\end{center}
\caption{Measurements of the bacterial hook bending stiffness $EI$.}
\label{table_EI}
\end{table}

\section{Effect of centrifugal force}
A simple calculation shows that the bending of the hook due to the centrifugal force during rotation is not sufficient to explain the observed movement of the bead towards the membrane during rotation. 
A bead of density $\rho_b$, radius $R_b$, and mass $m_b=\frac{4}{3}\pi R_b^3\rho_b$, rotating with an angular speed $\omega$ on a circular trajectory of radius $r$, is responsible for a centripetal and centrifugal force $F=m_b \omega^2 r$. 
Considering the limiting situation where the hook is a vertical solid cantilever anchored at one extremity, its stiffness is $k_{c}=3EI/L^3$, where $L$ is the lever arm, and the maximum deflection under the action of a constant force is $\delta_B = \frac{FL^3}{3EI}$.
Considering a latex bead ($\rho_b=1$ g/cm$^3$) with $R_b=0.5$ $\mu$m, $m_b=5\cdot 10^{-16}$ kg, rotating at $\omega=2\pi\cdot 50$ rad/s, on a trajectory of radius $r=200$ nm, the centrifugal force is of the order of $F\sim 10^{-17}$ N. Such force on a cantilever having the experimental value of the bending stiffness $EI=10^{-25}$ Nm$^2$ would induce a maximum deflection of only  $\delta_B\sim 10^{-12}$ m (considering the bead radius as lever arm).

\section{Comparison of hook bending stiffening under opposite twists}

\begin{figure}
\centering
\includegraphics[width=.99\linewidth]{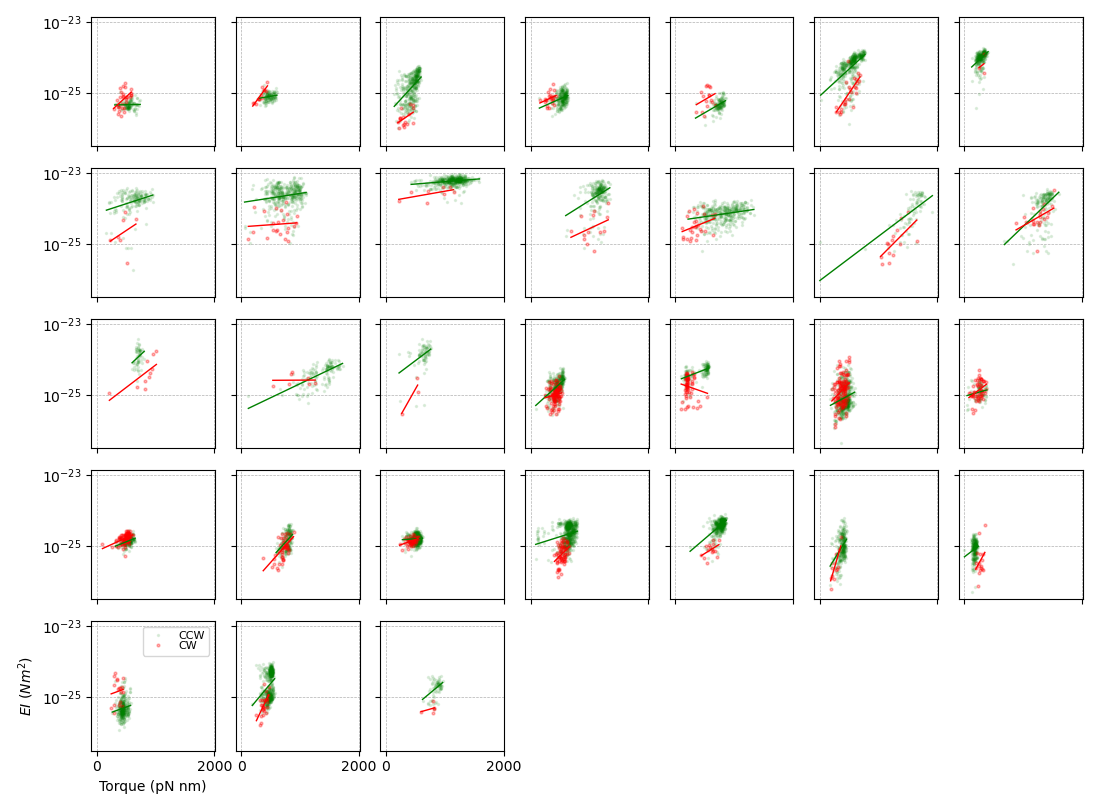}
\caption{Hook bending stiffness under opposite twists. The bending stiffness $EI$ is measured from resurrection traces of a switching strain, separating the hook response when the motor rotates CCW and CW. Each panel corresponds to a different resurrection trace of different motors rotating beads with $R_b=1000$ and 500 nm. Red: CW, green: CCW, linear fits are indicated by the lines.
}
\label{SIfig_EI_ccwcw}
\end{figure}

Given the heterogeneity of the mechanical responses of different hooks (fig.5 of the main text), we aimed at comparing the bending stiffness $EI$ of a single hook dynamically subject to opposite twists during the resurrection of the motor. This can be done by analyzing resurrection traces where the motor switches between CCW and CW rotation. To probe the entire response of $EI$ as a function of torque, we used large beads ($R_b=500, 1000$nm). Using the \textit{E. coli} parent strain where \textit{cheY} is not deleted resulted in resurrection traces where the motors switched direction most frequently at large stator number and high torque, while at low stator number and low torque very few switches to CW were observed. This prevented the quantification of $EI$ at low torque. With a $\Delta$\textit{cheRB} mutant we observed more frequent CW rotation intervals at low torque, and the results of these experiments are shown in fig. \ref{SIfig_EI_ccwcw}. 

The analysis of the signals described above, based on quantities extracted in time windows of a few seconds, works best when long dwell times with constant speed (either CCW or CW) are present in the trace. While this is possible in CCW rotation resurrections, we found that the dwell times in CW rotation were insufficiently long. This would produce artifacts, by mixing CW and CCW speed in single analysis time-windows (an example trace is shown in the last panel of fig.\ref{SIfig_EI_ccwcw}). We therefore proceed by locating all the CW dwell times, removing them from the trace and concatenating them at the end of the resulting trace, such that all the CW and CCW dwell times were grouped together. To avoid artifacts from analysing zero-speed regions and from the speed transitions between CCW and CW (of tens of ms time duration), we further remove from the analysis all points where the absolute speed $|\omega|$ is below a threshold of 1 Hz. The resulting trace can be analyzed as before, as the total time spent in the CW rotation is sufficient for the time-windowed analysis. 

These complications make the results shown in fig.\ref{SIfig_EI_ccwcw} less conclusive than those obtained with the non-switching strain. However, we can distinguish in a majority of cases that the bending stiffness $EI$ measured in CCW rotation (green points) is larger than in the CW rotation (red points). Yet, we sometimes see the inverse, or cases where the EI values are very similar. This qualitatively suggests that  an asymmetry may be present in the way the hook stiffens when twisted in the two directions, but more data would be required to provide a clear and quantitative answer.

\subsection{cheRB mutant preparation}
The MT02 strain was used as the recipient for inactivation of the \textit{cheRB} genes using $\lambda-$red mediated recombination method as described in \cite{wanner2000one}. Briefly, the chloramphenicol resistance gene (\textit{cat}) was amplified from the pWRG100 plasmid \cite{blank2011rapid} using primers tapfwd and cheYrev, which imparted flanking homologous regions upstream and downstream of the \textit{cheRB} genes. The MT02 strain was first transformed with the pKD46 plasmid encoding the Red recombinase and electroporated with the amplified PCR fragment. The chloramphenicol-resistant recombinant clones were purified twice on LB plates supplemented with chloramphenicol (25 mg/L) and characterized by PCR using primers cheRBfwd and cheRBrev. Phage P1 was used to transduce the mutation into the parental MT02 strain, yielding \textit{cheRB::cat} MT02 strain. The oligonucleotides used in this study are described in Table \ref{table_olig}.

\begin{table}[t]
\begin{center}
\begin{tabular}{l l}
Name & Primer sequence (5’ - 3’) \\ 
\hline
\textit{tapfwd} & TGCAGTTACAAATTGCGCCAGTGGTATCCTGAAGT \\
       & GATTGAGAAGGCGCTCGCCTTACGCCCCGCCCTGC \\
\textit{cheYrev} & AACCAAAAATTTAAGTTCTTTATCCGCCATTTCA \\
        & CACTCCTGATTTAAATCTAGACTATATTACCCTGTT \\
\textit{cheRBfwd} & GTCGCGTGTGGCGGTATTTACCC \\
\textit{cheRBrev} & CCGCCTGCCTGCAACTTATTGAGAG \\
\hline
\end{tabular}
\end{center}
\caption{Oligonucleotide list.}
\label{table_olig}
\end{table}

\printbibliography

\end{document}